\documentclass[10pt,twocolumn,final,3p]{elsarticle}

\usepackage{amsmath}
\usepackage[]{graphicx}
\usepackage{subfigure}
\usepackage[utf8]{inputenc}
\usepackage[pdftex]{hyperref}
\usepackage{xspace} 

\newcommand{\etal}{\textit{et al}.\@\xspace}
\newcommand{\ie}{\textit{i.e.}\@\xspace}
\newcommand{\cf}{\textit{cf}.\@\xspace}
\newcommand{\siesta}{\textsc{Siesta}\@\xspace}
\newcommand{\pwscf}{\textsc{Pwscf}\@\xspace}
\newcommand{\vasp}{\textsc{Vasp}\@\xspace}
\newcommand{\abinitio}{\textit{ab initio}\@\xspace}
\newcommand{\Abinitio}{\textit{Ab initio}\@\xspace}

\newcommand{\ud}{\mathrm{d}}

\newcommand{\sigmaB}{\sigma_{0001}}
\newcommand{\sigmaP}{\sigma_{10\bar{1}0}}
\newcommand{\sigmaPi}{\sigma_{10\bar{1}1}}

\journal{Acta Materialia}

\begin{document}

\begin{frontmatter}

	\title{Vacancy clustering in zirconium: an atomic scale study\tnoteref{t1}}
	\tnotetext[t1]{Article published in Acta Mater. \textbf{78}, 65--77 (2014).\\
	\url{http://dx.doi.org/10.1016/j.actamat.2014.06.012}}

	\author{Céline Varvenne\fnref{1}}
	\fntext[1]{Present address: Institute of Mechanical Engineering, 
	École Polytechnique Fédérale de Lausanne, 
	Lausanne CH-1015, Switzerland}

	\author{Olivier Mackain}

	\author{Emmanuel Clouet\corref{1}}
	\ead{emmanuel.clouet@cea.fr}
	\cortext[1]{Corresponding author}

	\address{CEA, DEN, Service de Recherches de Métallurgie Physique, F-91191 Gif-sur-Yvette, France}

\begin{abstract}
The stability properties of vacancy clusters in hexagonal close-packed Zr, cavities and dislocation loops, are investigated at the atomic scale, with a modeling approach based on density functional theory and empirical potentials. 
Considering the vacancy-vacancy interactions and the stability of small vacancy clusters, we establish how to build the larger clusters. 
The study of extended vacancy clusters is then performed using continuous laws for defect energetics. Once validated with an empirical potential, these laws are parameterized with \textit{ab initio} data. 
Our work shows that the easy formation of $\langle a \rangle$ loops
can be explained by their thermodynamic properties.
\end{abstract}


\end{frontmatter}

\section{Introduction}

Zirconium alloys are widely used in the nuclear industry as a cladding material.
In nuclear reactors, they are subjected to a fast neutron flux,
leading to the creation of a large amount of point defects, both vacancies and self-interstitials.
These point defects then diffuse and can be trapped by the different sinks of the system,
or can cluster to form larger defects, like dislocation loops and cavities \cite{Onimus2012}.
Vacancy clusters can also appear in quenched zirconium alloys \cite{Carpenter1973}.

Extensive experimental studies have been carried out in the past to determine the structure of these defect clusters
in hexagonal close-packed (hcp) Zr and its alloys (see Ref.~\cite{Onimus2012} for a recent review).
At low irradiation doses,	
perfect dislocation loops with $\langle a \rangle=1/3\,\langle 11\bar{2}0 \rangle$ Burgers vector are observed
\cite{Northwood1979,Griffiths1987b,Griffiths1988}.
These $\langle a \rangle$ loops are both of interstitial and of vacancy type. 
Their habit plane is close to the prismatic plane of the hcp lattice.
The same perfect loops, all of vacancy type, are obtained in quenched Zr alloys \cite{Carpenter1973}.
Under irradiation, loops lying in the basal planes are also observed for the highest irradiation doses \cite{Griffiths1987b}. 
These loops are faulted with a Burgers vector $1/6\,\langle20\bar{2}3\rangle$, thus with a $\langle c \rangle$ component.
They are all of vacancy type.
Finally, cavities are observed in only a very few specific cases \cite{Griffiths1993b,Griffiths1994}. 

These vacancy and interstitial clusters have important consequences 
on the macroscopic behavior of zirconium.
Like in other metals, strong hardening is associated with the presence of these defects \cite{Onimus2012,Boyne2013}.
Irradiation also leads in hcp Zr to dimensional changes without any applied stress:
a Zr single crystal undergoes under irradiation an elongation along the $\langle a \rangle$ axis
of the hcp lattice and a shortening along the $\langle c \rangle$ axis,
with no significant volume change \cite{Carpenter1988}. 
The growth strain remains small at low fluence,
but a breakaway growth is observed at higher fluence \cite{Carpenter1988,Fidleris1988}.
This breakaway is correlated with the appearance of the $\langle c \rangle$ type vacancy loops
\cite{Griffiths1987b,Griffiths1987a}.

Understanding the formation of these clusters is of prime importance to be able 
to model the kinetic evolution of the microstructure under irradiation 
and of the associated macroscopic behavior. 
This requires first to know the relative stability of these clusters,
in particular vacancy clusters for which different types coexist.
Atomistic simulations appear as the suitable tool for such a study,
as they can provide information on cluster sizes which are not accessible by other techniques
and which are necessary to build higher level models or theories.
Several works already attempted to address this question \cite{Kapinos1992,Kulikov2005,Diego2008,Diego2011}. 
They either showed that the most stable vacancy clusters are cavities \cite{Kulikov2005}
or basal loops \cite{Kapinos1992,Diego2011}. 
This appears in contradiction with experimental observation
indicating that the easiest clusters to form are vacancy loops lying in the prismatic planes.
But all these simulations relied on empirical potentials, 
either long ranged pair potential \cite{Kapinos1992}
or Embedded Atom Method (EAM) \cite{Kulikov2005,Diego2008,Diego2011}.
These central forces empirical potentials are known to poorly model stacking faults
in hcp transition metals. 
Legrand \cite{Legrand1984} showed that one needs to correctly account for the electronic 
filling of the valence d band, and thus to consider the angular dependence 
of the atomic bonding, in order to obtain a good description of these stacking faults.
As vacancy loops, at least the smallest ones, are faulted,
it is worth looking at the stability of the vacancy clusters in hcp Zr 
with a better modeling of the atomic bonding than with the previously used 
empirical potentials. 
\Abinitio calculations represent a nice alternative but they 
can only be used to study small clusters containing a few vacancies.
We therefore propose to use an hybrid approach relying both on \abinitio calculations 
and empirical potentials to model these clusters. 

In this work we focus on the stability of vacancy clusters in hcp Zr.
Stability of small clusters, as well as stacking-fault and surface energies, 
are investigated with \abinitio calculations 
and then compared to predictions obtained with two recent EAM potentials developed by Mendelev and Ackland \cite{Mendelev2007}.
As a result of this comparison, one EAM potential is selected to study larger vacancy clusters. 
This allows us to validate analytical laws based on continuous models able to describe 
their formation energies.
These analytical laws are finally parameterized on  \abinitio calculations 
so as to conclude on the relative stability of the different vacancy clusters in hcp Zr.

\section{Details of atomistic simulations}

Our \abinitio calculations  are based on the Density Functional Theory (DFT), using the \pwscf code of the Quantum Espresso package \cite{Giannozzi2009}. All calculations  are performed in the Generalized Gradient Approximation with the exchange-correlation functional of Perdew-Burke-Ernzerhof \cite{Perdew1996}. Valence electrons are described with plane waves, using a cutoff of $28$\,Ry. The core electrons are replaced by an ultrasoft pseudo-potential of Vanderbilt type, including 4s and 4p electrons as semicore. The electronic density of state is broadened with the Methfessel-Paxton function, with a broadening of $0.3$\,eV. The integration is performed on a regular grid of $14 \times 14 \times 8$ k-points for the primitive cell and an equivalent density of k-points for the supercells used in defect calculations. 
This \abinitio modeling approach has been already validated on Zr bulk properties in a previous study \cite{Clouet2012}.

\Abinitio calculations of vacancy clusters, including the single vacancy, are performed in a periodic supercell 
corresponding to $5 \times 5 \times 4$ hcp primitive unit cells and containing 200 atomic lattice sites. 
Only the atomic positions are relaxed while the periodicity vectors are kept fixed
(constant volume calculations).
The elastic correction described in Ref. \cite{Varvenne2013} is applied 
so as to remove the elastic interaction of the vacancy cluster 
with its periodic images.

The two EAM potentials we used were developed by Mendelev and Ackland \cite{Mendelev2007}. 
They are labeled $\#2$ and $\#3$ in Ref.~\cite{Mendelev2007}.
Both of them give a reasonable description of the bulk properties of hcp Zr.
EAM $\#3$ potential has already been used to calculate the properties of small vacancy and interstitial clusters by De Diego \etal \cite{Diego2011}.
It is believed to be better to study defect properties in hcp Zr,
as some stacking fault energies in the basal and prism planes have been adjusted on \abinitio values.
EAM $\#2$ potential is particularly designed to describe the hcp-bcc transition, 
but it also gives a reasonable description of defects in hcp Zr.
We will see in the following that it is actually better suited than the EAM $\#3$ potential 
to study vacancy clustering.
Atomistic simulations with these empirical potentials are performed with a $100 \times 100 \times 50$ supercell 
containing 1 million of atomic lattice sites. This size gives well converged energies for all the investigated defects. 

\begin{table*}[htbp]
\caption{Vacancy properties in hcp Zr: formation energies $E^{\rm for}$,
migration energies $E^{\rm mig}_{\rm bas}$ and $E^{\rm mig}_{c}$, respectively in the basal plane
and along the $\langle c \rangle$ axis,
non-null components of the elastic dipole tensor,
and relaxation volumes. 
Energies and dipoles are given in eV  and the relaxation volume is normalized by the atomic volume $\Omega$.}
\label{Tab:V_Zr_pwscf}
\begin{center}
\begin{tabular}{lcccccc}
\hline \hline
		& $E^{\rm for}$  & $E^{\rm mig}_{\rm bas}$ & $E^{\rm mig}_{c}$ & $P_{11}=P_{22}$ & $P_{33}$ & $\delta V^{\rm rel}$\\ 
		& (eV)		 & (eV)			   & (eV)	       & (eV)		 & (eV)	    & ($\Omega$)\\
  \hline
 \Abinitio (this work)
 		& 2.07 & 0.54 & 0.65 & $-4.90$ & $-7.06$ & $-0.40$\\
 \Abinitio (\siesta \cite{Verite2007a}) 
 		& 2.14 & 0.55 & 0.66  				\\ 
 EAM $\#2$ 	& 2.26 & 1.03 & 1.12 & $-0.63$ & $-0.78$ & $-0.05$	\\ 
 EAM $\#3$ 	& 1.67 & 0.63 & 0.72 & $-5.55$ & $-5.55$ & $-0.38$	\\ 
 Expt. (resistivity) \cite{Neely1970} 		&    		&  \multicolumn{2}{c}{$0.58\pm 0.04$} \\ %
 Expt. (PAS) \cite{Hood1984,Hood1986}	&   $\geq 1.5$	& \multicolumn{2}{c}{$0.65\pm 0.05$} 	 \\ %
 Expt. (growth kinetics) \cite{Buckley1980}	&	& \multicolumn{2}{c}{0.65} \\
 Expt. (Huang) \cite{Ehrhart1982,Ehrhart1986}  	&	&	&	&&&  $-0.1$ \\
\hline \hline
\end{tabular}
\end{center}
\end{table*}

Before using these different atomic models to study vacancy clustering, 
it is worth comparing their results for the single vacancy properties.
The obtained vacancy formation and migration energies are compared with experimental 
data in Table~\ref{Tab:V_Zr_pwscf}.
All  three models lead to a vacancy formation energy which is compatible with the lower-bound value given by positron annihilation spectroscopy (PAS) \cite{Hood1986}.
The vacancy can migrate along two non-equivalent pathways:
one inside the basal plane ($E^{\rm mig}_{\rm bas}$) and the other one out of the basal plane ($E^{\rm mig}_{c}$). 
The migration energies obtained with \abinitio calculations are $0.54$~eV in the basal plane and $0.65$~eV out of the basal plane.
This suggests a significant anisotropy of vacancy diffusion, with a fast diffusion inside the basal plane. 
Previous \abinitio calculations in Zr \cite{Verite2007a} have already shown such an anisotropy,
and the same anisotropy is obtained with the EAM potentials.
This is in agreement with the experimental characterization of self-diffusion performed by Hood \etal \cite{Hood1995,Hood1997},
who obtained a ratio of $0.6\pm0.2$ between the diffusion coefficient along the $\langle c \rangle$ axis and in the basal plane.
In addition, the average migration energy given by \abinitio calculations is in very good agreement with the experimental ones,
deduced either from resistivity recovery \cite{Neely1970}, 
positron annihilation spectroscopy (PAS) \cite{Hood1984,Hood1986}, 
or TEM characterization of irradiation growth \cite{Buckley1980}.
Whereas the average migration energy is also good with EAM $\#3$, EAM $\#2$ overestimates this energy.

We also use the methodology of Ref.~\cite{Varvenne2013}
to deduce from our atomistic simulations the elastic dipole when  the vacancy is
in its stable configuration.
The \abinitio values (Tab.~\ref{Tab:V_Zr_pwscf}) indicates a contraction more important 
along the $\langle c \rangle$ axis than in the basal plane. 
This is consistent with the anisotropic displacement field 
evidenced by Ehrhart \etal \cite{Ehrhart1982, Ehrhart1986}, 
using Huang X-ray diffuse scattering experiments.
The relaxation volume of the vacancy, $\delta V^{\rm rel}$, can be deduced 
from this elastic dipole.\footnote{$\delta V^{\rm rel} =
\frac{ (C_{33}-C_{13})(P_{11}+P_{22}) + (C_{11}+C_{12}-2C_{13})P_{33} }{ (C_{11}+C_{12})C_{33} - 2 {C_{13}}^2 }$,
where $C_{ij}$ are the elastic constants of hcp Zr.}
\Abinitio calculations lead to a higher relaxation volume than the experimental value
reported by Ehrhart \etal \cite{Ehrhart1982,Ehrhart1986}.
The relaxation volume given by EAM \#3 potential is in good agreement with \abinitio results, 
whereas EAM \#2 potential leads to a very small relaxation volume.

\section{Stability of small vacancy clusters}
\label{sec:smallClusters}

The stability of small vacancy clusters is studied by calculating 
their binding energies. 
After building the configurations corresponding to the chosen vacancy clusters,
atomic positions are relaxed with a conjugate gradient algorithm. 
We define then the binding energy of a cluster containing $n$ vacancies
as the difference between the formation energies of $n$ isolated vacancies 
and the formation energy of the cluster: 
\begin{align*}
	E^{\rm b}(\textrm{V}_n) = & n E^{\rm f}(\textrm{V}_1) - E^{\rm f}(\textrm{V}_n) \\
	= & n E(\textrm{V}_1) - E(\textrm{V}_n) - (n-1) E(\textrm{bulk}),
\end{align*}
where $E(\textrm{V}_1)$, $E(\textrm{V}_n)$ and $E(\textrm{bulk})$
are the energies of the same simulation cell containing respectively one isolated vacancy, 
the vacancy cluster and no defect.
A positive value of the binding energy indicates that the interaction between the vacancies 
is attractive and that the cluster is stable.

\subsection{Divacancy}

\begin{figure}[!bth]
	\begin{center}
		\includegraphics[width=0.9\linewidth]{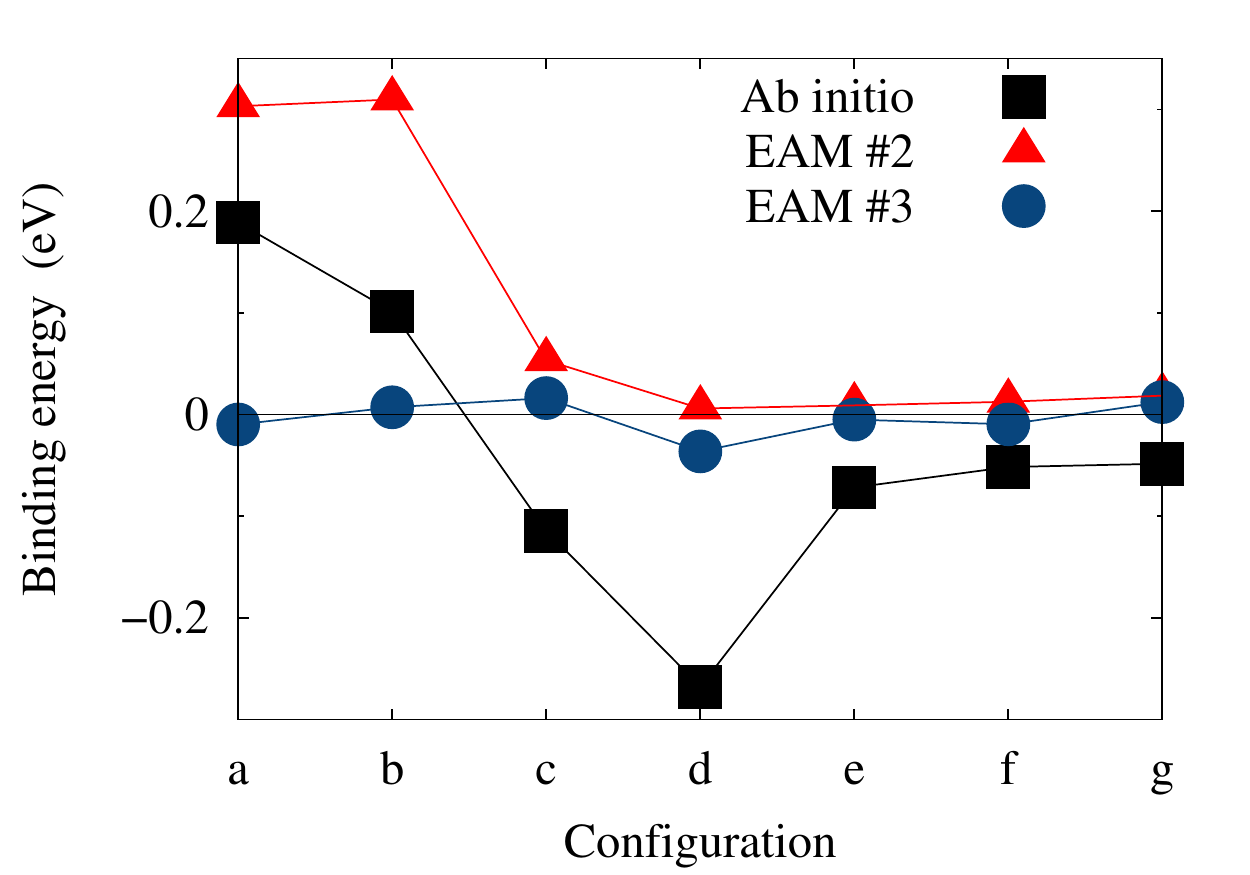}
		\includegraphics[width=0.5\linewidth]{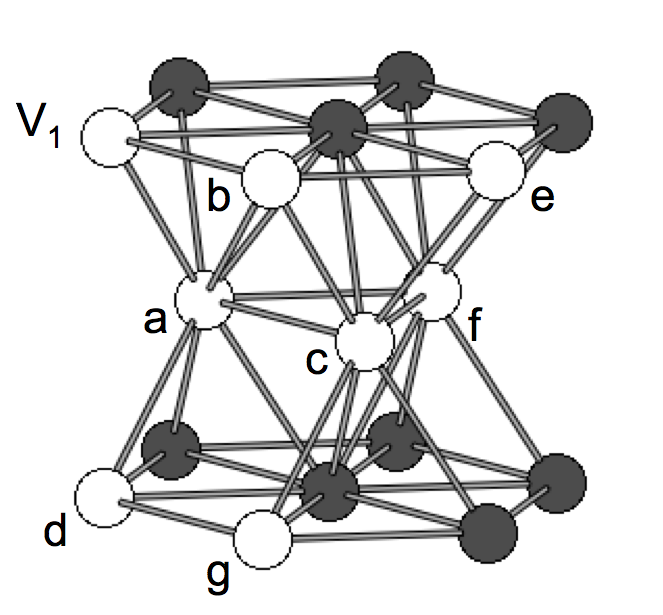}
	\end{center}
	\caption{Binding energies of a divacancy 
	calculated with the different energy models.
	The corresponding divacancy configurations 
	are sketched on the hcp lattice, where
	$\textrm{V}_1$ denotes the position of the first vacancy and the letters 
	a to g the position of the second one.}
	\label{fig:El_VV_dist}
\end{figure}

In order to understand how to build the vacancy clusters, we first analyse the interaction between two vacancies. 
Different configurations are investigated: 
the second vacancy is placed on the successive neighboring shells of the first vacancy, 
at distances lower than two lattice parameters 
(see Fig.~\ref{fig:El_VV_dist} for the detailed configurations). 
The corresponding binding energies are presented in Fig.~\ref{fig:El_VV_dist}.

\Abinitio calculations show that the interaction is attractive only when the vacancies are first nearest neighbors.
The first nearest-neighbor configuration with the divacancy lying in the $1/6\ [02\bar{2}3]$ direction (a configuration) 
is twice more attractive than the b configuration lying in the basal plane.
All other configurations, corresponding to vacancies separated by more than one first nearest-neighbor distance, are repulsive.
In particular, the d configuration, which lies along the $\langle c \rangle$ axis, shows a strongly negative binding energy ($-0.26$~eV). 
For the most distant e to g configurations, the magnitude of the binding energy decreases, 
but the interaction still remains repulsive.
Similar results are found for divacancies in hcp titanium \cite{Raji2009, Connetable2011}: 
only first nearest-neighbor configurations are attractive, 
with however almost equal values for both a and b configurations, 
and the d configuration lying along the $\langle c \rangle$ axis is strongly repulsive.

The same divacancy configurations are investigated with the EAM potentials $\#2$ and $\#3$. 
The binding energies obtained with EAM $\#3$ are close to zero for all the configurations (Fig.~\ref{fig:El_VV_dist}). 
The vacancies do not interact, even when the vacancies are first nearest-neighbors.
This is incompatible with the DFT results and with the vacancy clustering observed in experiments. 
This potential is therefore not well suited to describe vacancy clustering in hcp Zr, and we will mainly ignore it in the following.
On the other hand, EAM $\#2$ potential leads to an attractive interaction when vacancies are first nearest-neighbors,
and to zero binding when the vacancies are further. 
This is qualitatively consistent with \abinitio results, although the empirical potential overestimates the divacancy stability and does not account for the repulsive character of the c to g configurations.
It is also worth pointing out that the potential fails to discriminate between the a and b configurations. 
This is a direct consequence of the central force approximation used by EAM potentials, 
where no angular dependence is included.
As a consequence, these potentials cannot catch the difference between configurations a and b where 
the two vacancies composing the divacancy have the same environment 
and are separated by an almost equal distance.
The same limitation of EAM potentials will impact the relative stability of vacancy clusters 
predicted by EAM \#2 potential, as it will be seen later.

\subsection{Compactness of stable clusters: the tri-vacancy}

\begin{table}[!bhtp]
	\caption{Total binding energy for different configurations of a tri-vacancy (in eV). 
		The vacancies composing the cluster are sketched by squares 
		on a basal projection of the hcp lattice, whereas spheres correspond to atoms. 
		White and grey symbols are lying respectively in the 
		$z=0$ and $z=c/2$ basal planes. 
		When necessary, squares of different sizes are used for vacancies
		lying in different basal planes separated by a distance $c$.}
	\label{Tab:additiv_VV}
	\centering
	\begin{tabular}{llcc}
		\hline\hline
		\multicolumn{2}{c}{Configuration} & \Abinitio & EAM \#2 \\
		\hline
		1 & \includegraphics[scale=0.15]{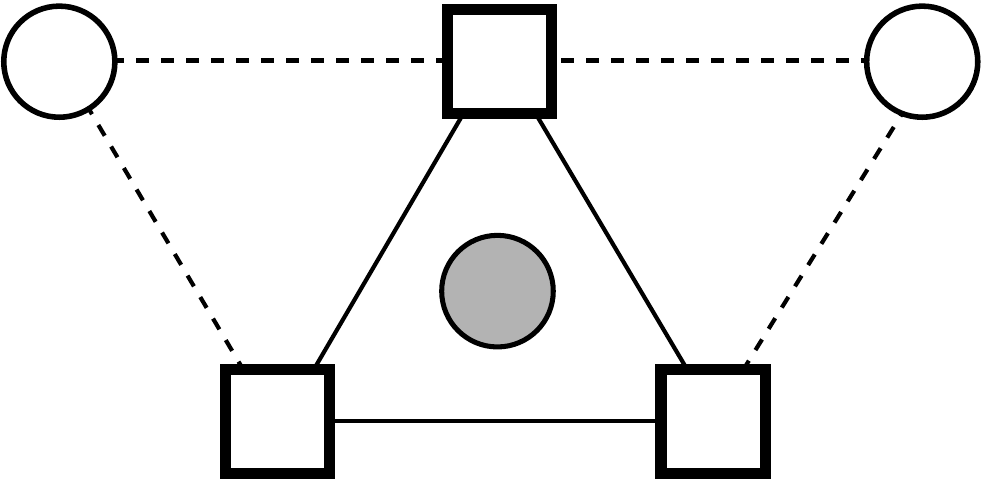}		& 0.40	& 0.91	\\  
		2 & \includegraphics[scale=0.15]{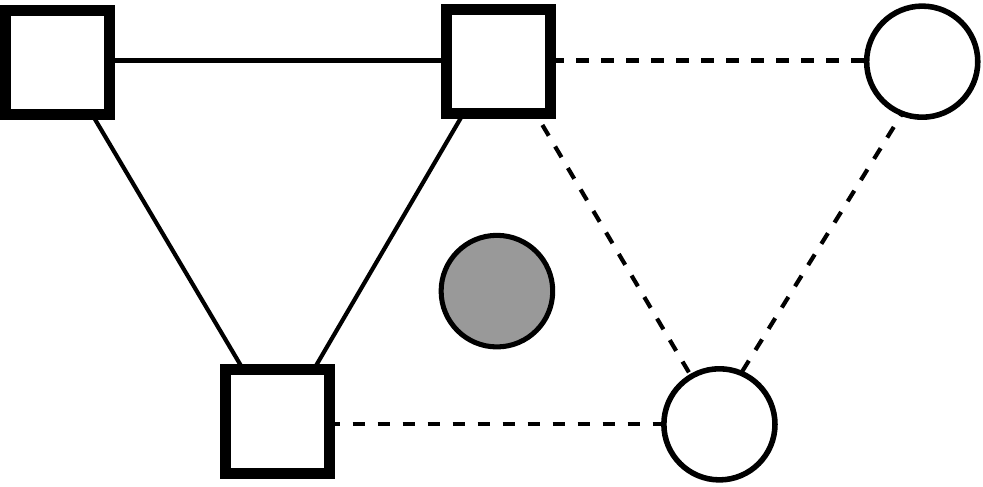}	& 0.55	& 1.02	\\  
		3 & \includegraphics[scale=0.15]{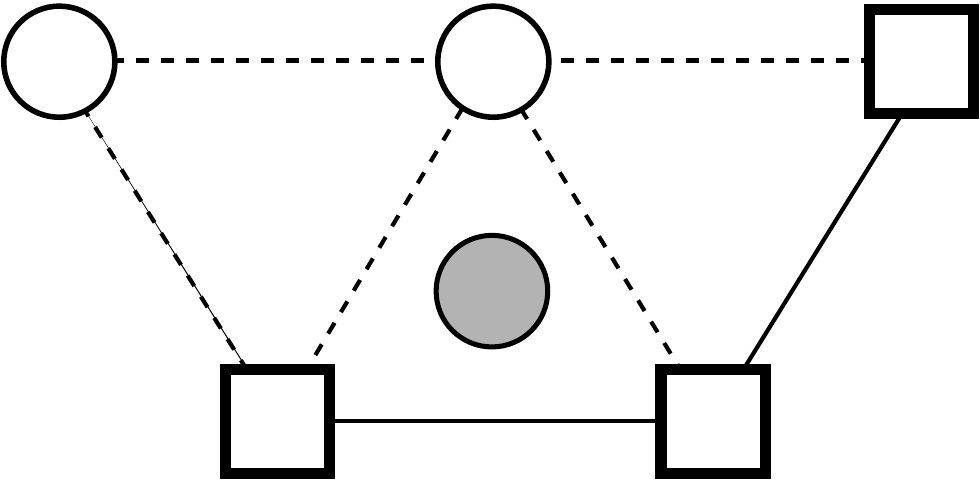}		& 0.17	& 0.74	\\  
		4 & \includegraphics[scale=0.15]{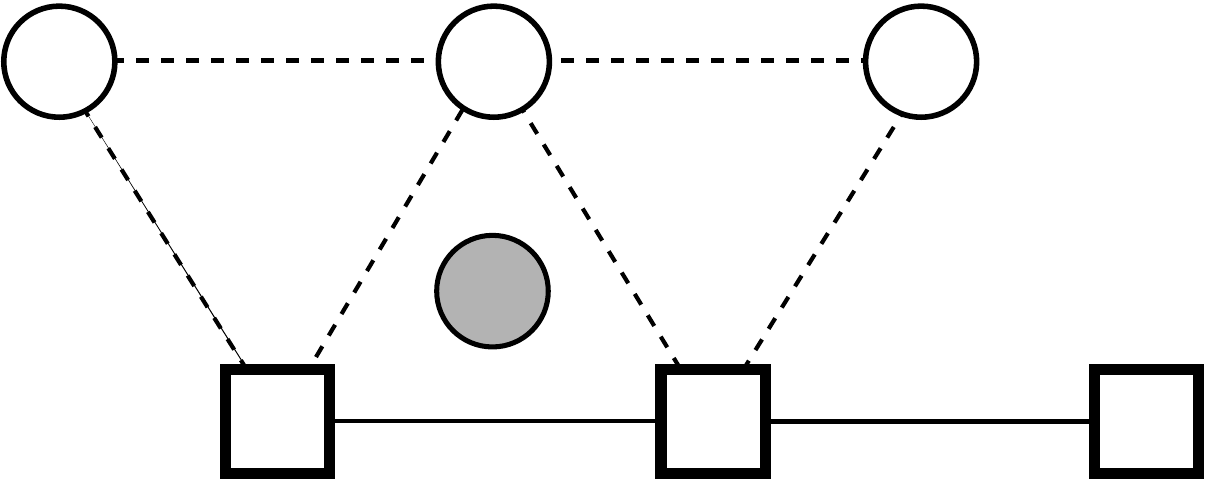}		& 0.23	& 0.76	\\  
		5 & \includegraphics[scale=0.15]{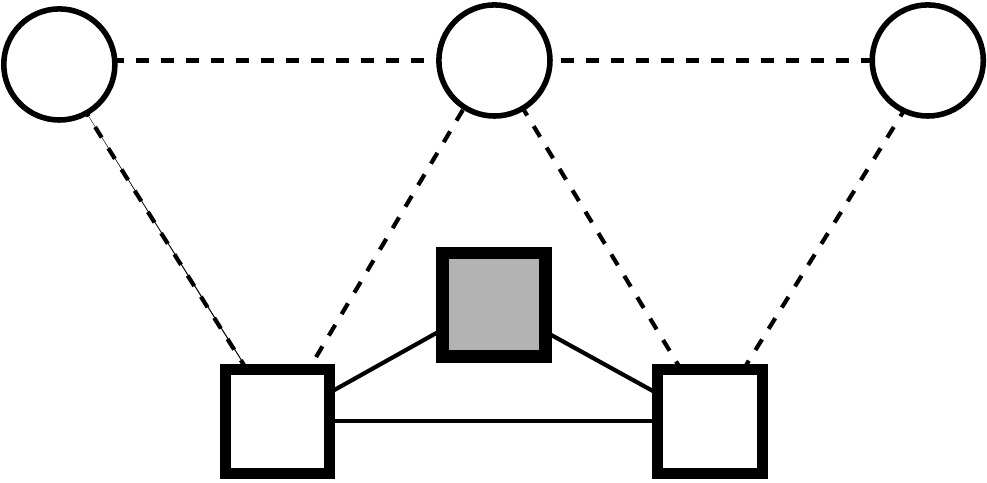}		& 0.53	& 0.89	\\  
		6 & \includegraphics[scale=0.15]{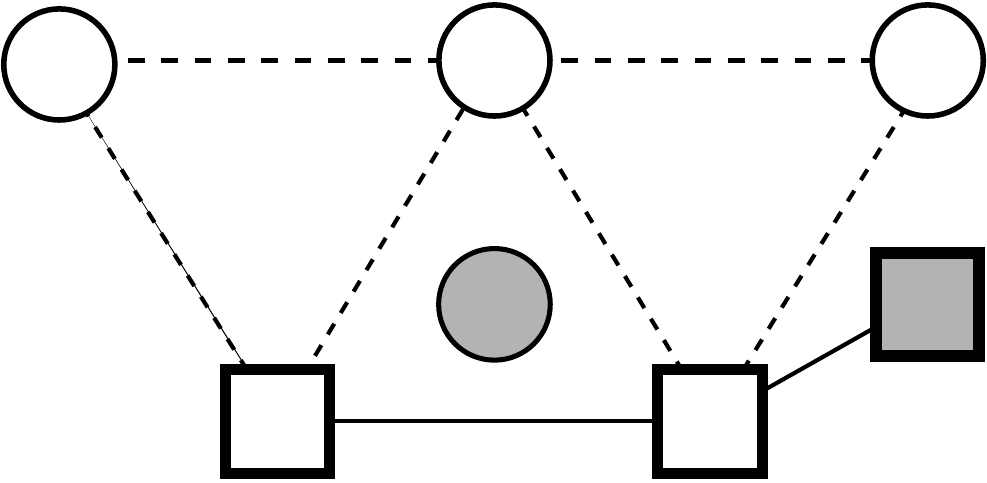}		& 0.27	& 0.74	\\  
		7 & \includegraphics[scale=0.15]{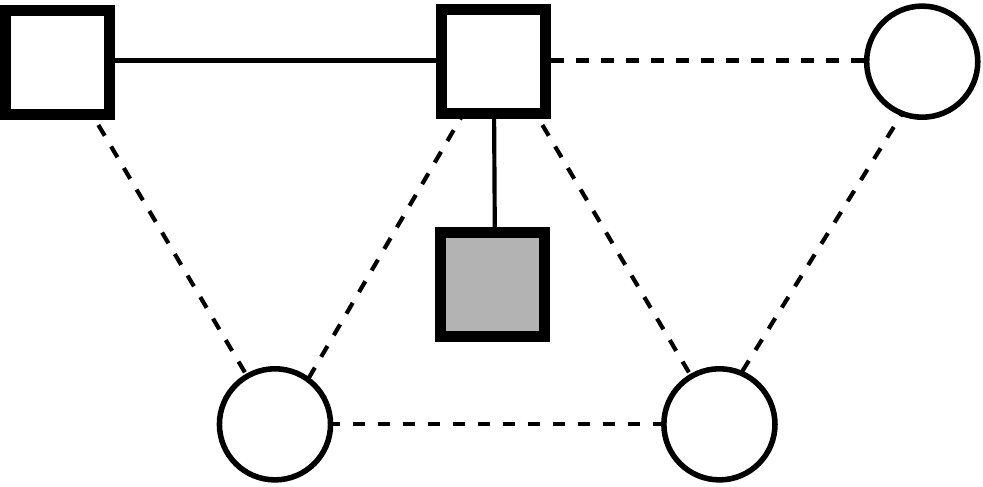}		& 0.33 	& 0.78	\\  
		8 & \includegraphics[scale=0.15]{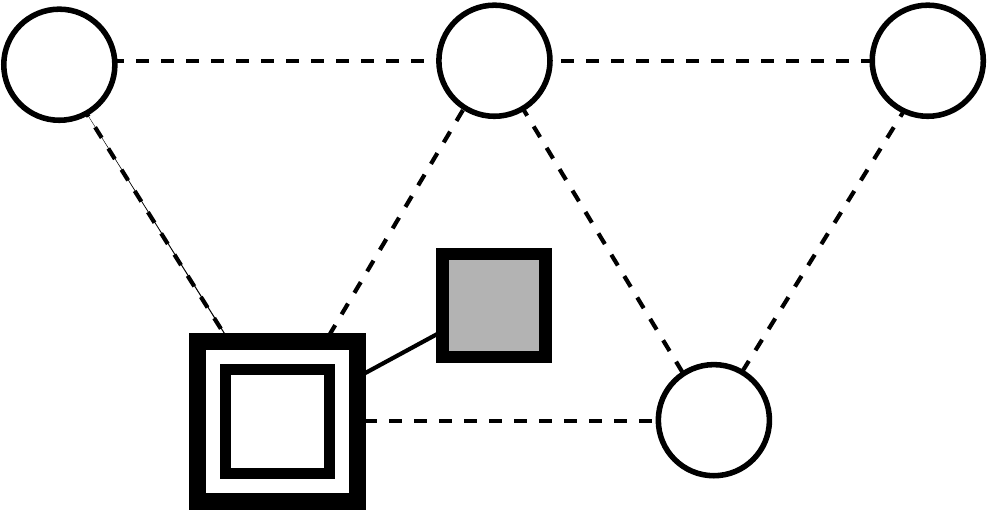}	& 0.20 	& 0.72	\\  
		9 & \includegraphics[scale=0.15]{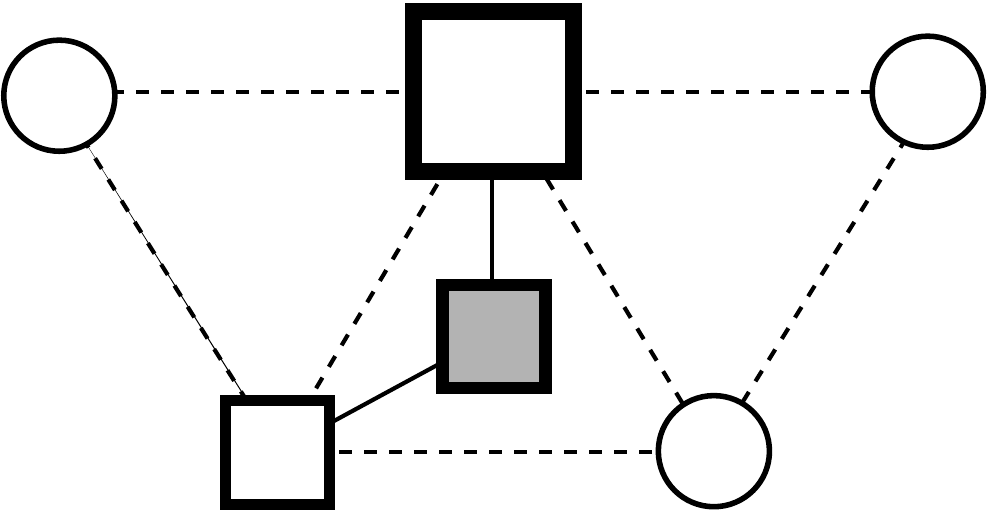}	& 0.34 	& 0.76	\\  
		\hline\hline
	\end{tabular}
\end{table}

We now look at how to build larger vacancy clusters. 
Based on the results obtained for the divacancy, 
only clusters formed by vacancies which are first nearest-neighbors are considered.
This leads for the tri-vacancy to nine different clusters (Table \ref{Tab:additiv_VV}).
Both \abinitio calculations and the EAM \#2 potential predict positive binding energies
for all these nine clusters.
The most stable clusters are the compact ones, which involve the largest number of first nearest-neighbor interactions
(clusters 1, 2 and 5).
Two different compact configurations can be formed in a basal plane, which are crystallographically not equivalent
(clusters 1 and 2).
With both interaction models, the most stable one is the configuration 2.
The compact tri-vacancy lying in a prismatic plane (configuration 5) 
has the same formation energy as the most stable basal configuration.

\subsection{Relative stability of compact clusters}

\begin{figure}[!bthp]
	\begin{center}
		\begin{tabular}{lccc}
			\hline\hline
			& basal	& prismatic	& 3D	\\
			\hline
		V$_3$ &
		\includegraphics[scale=0.16]{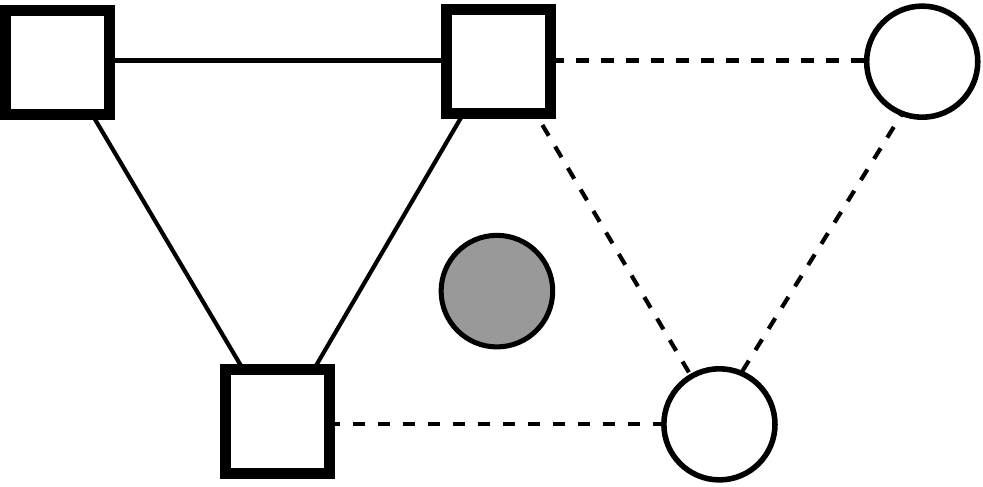} 	&
		\includegraphics[scale=0.16]{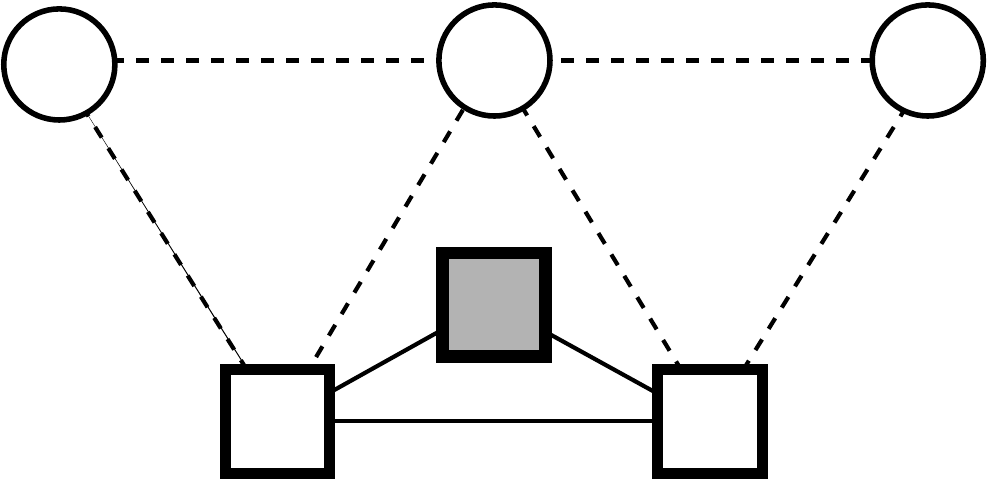}&
		\\
		V$_4$ &
		\includegraphics[scale=0.16]{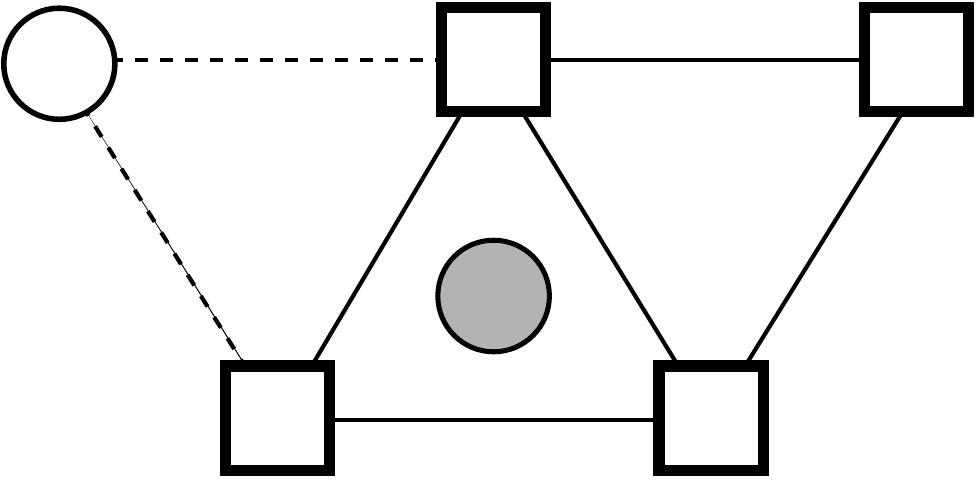} 	&
		\includegraphics[scale=0.16]{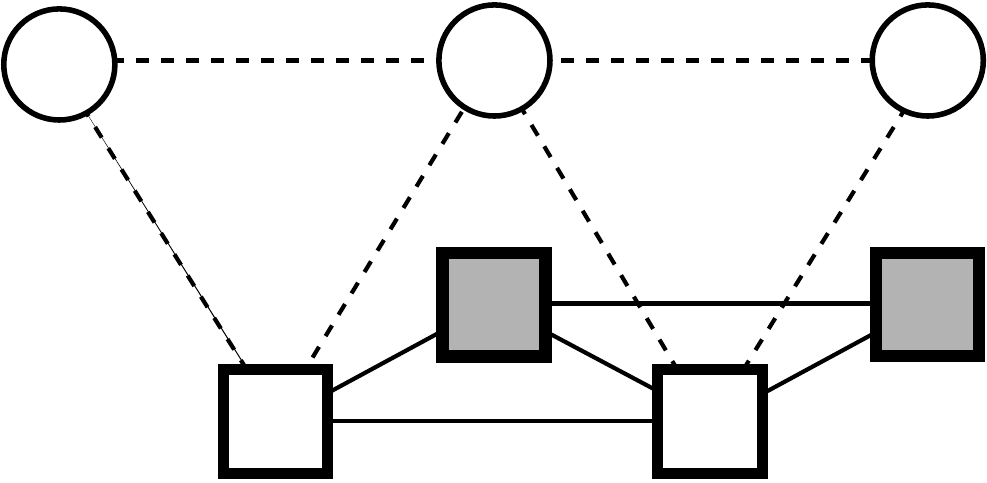}&
		\includegraphics[scale=0.16]{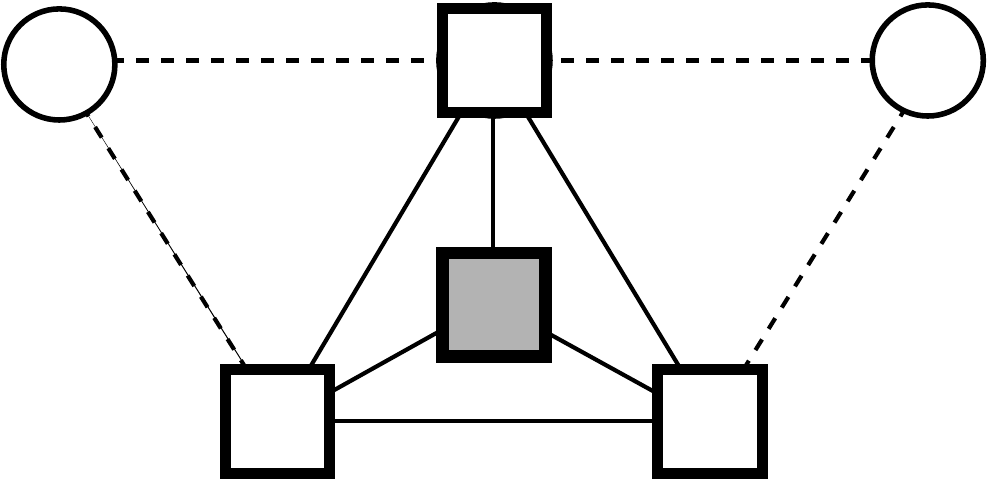}
		\\
		V$_5$ &
		\includegraphics[scale=0.16]{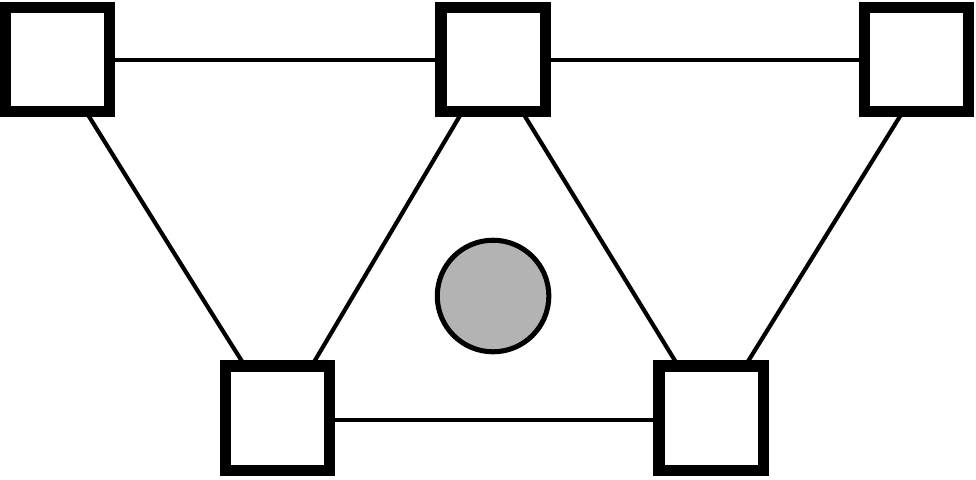} 	&
		\includegraphics[scale=0.16]{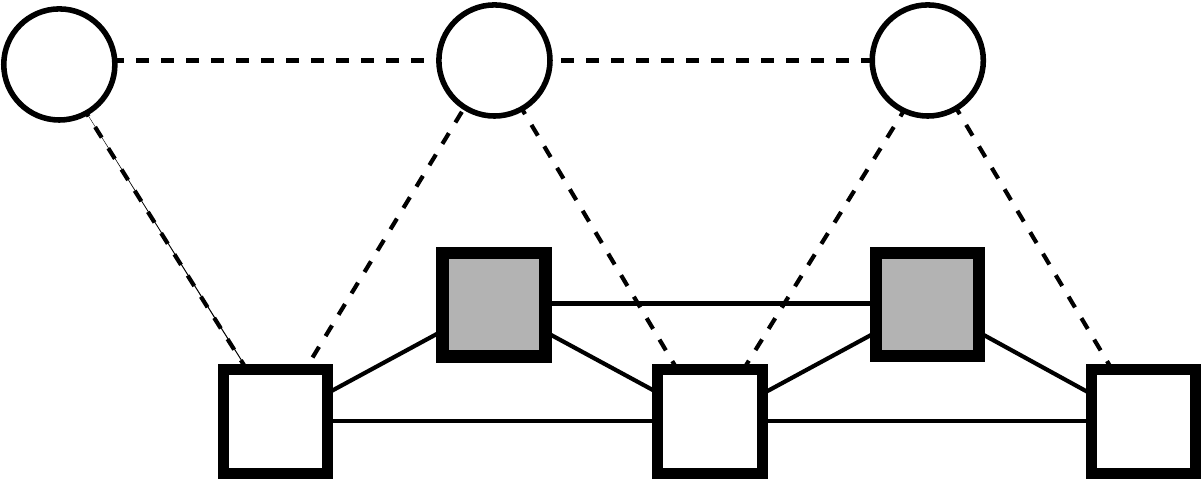}&
		\includegraphics[scale=0.16]{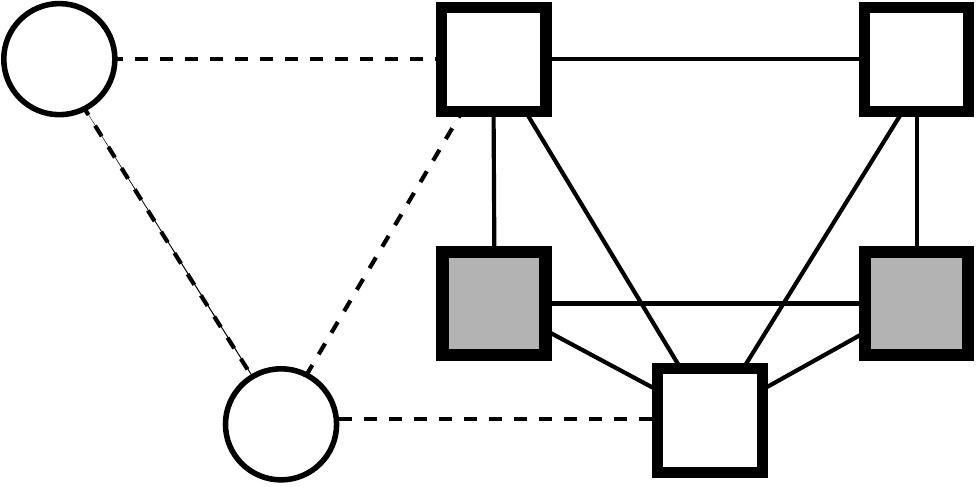}	
		\\
		V$_6$ &
		\includegraphics[scale=0.16]{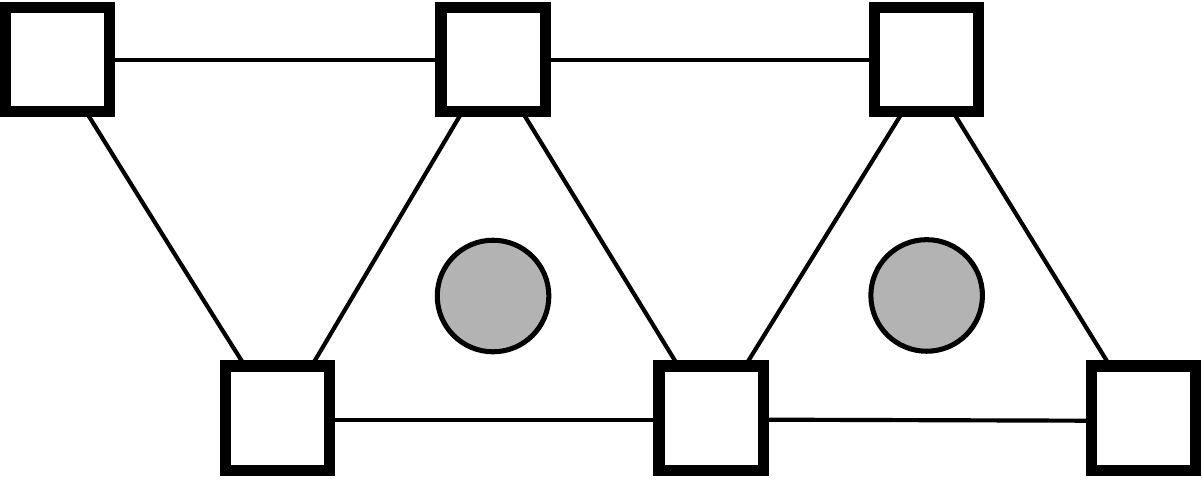} 	&
		\includegraphics[scale=0.16]{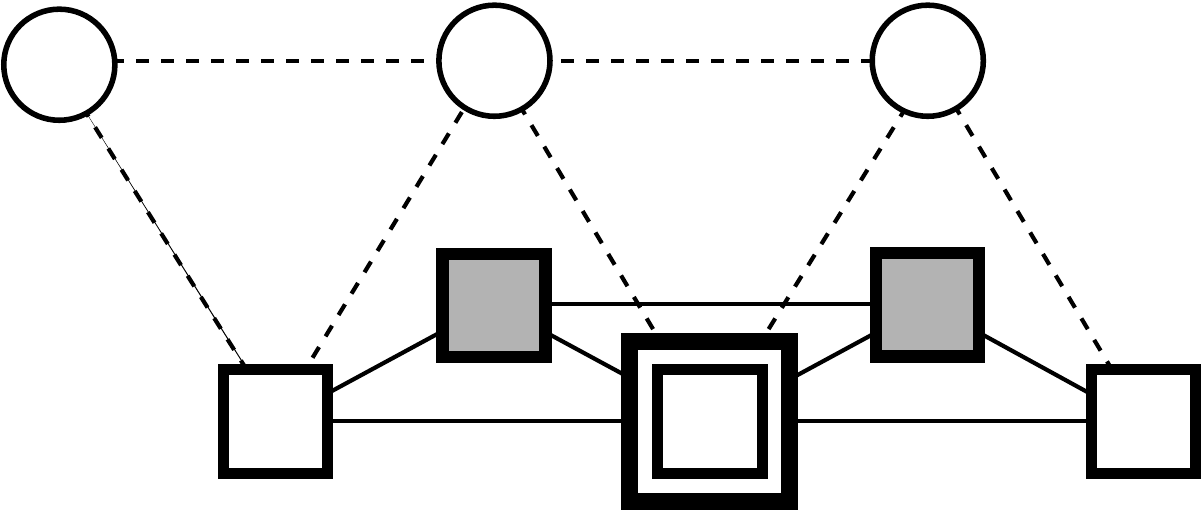}&
		\includegraphics[scale=0.16]{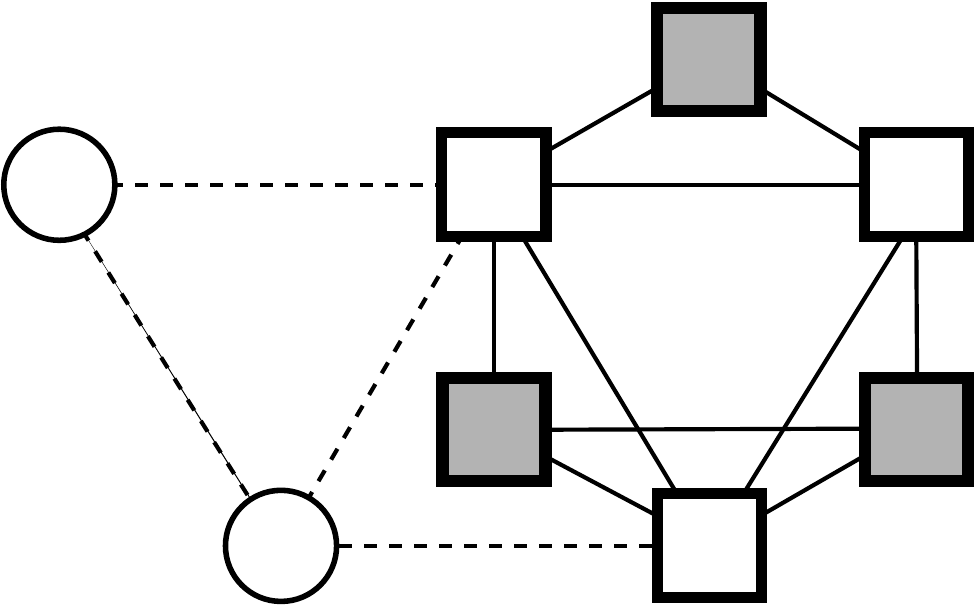}	
		\\
		V$_7$ &
		\includegraphics[scale=0.16]{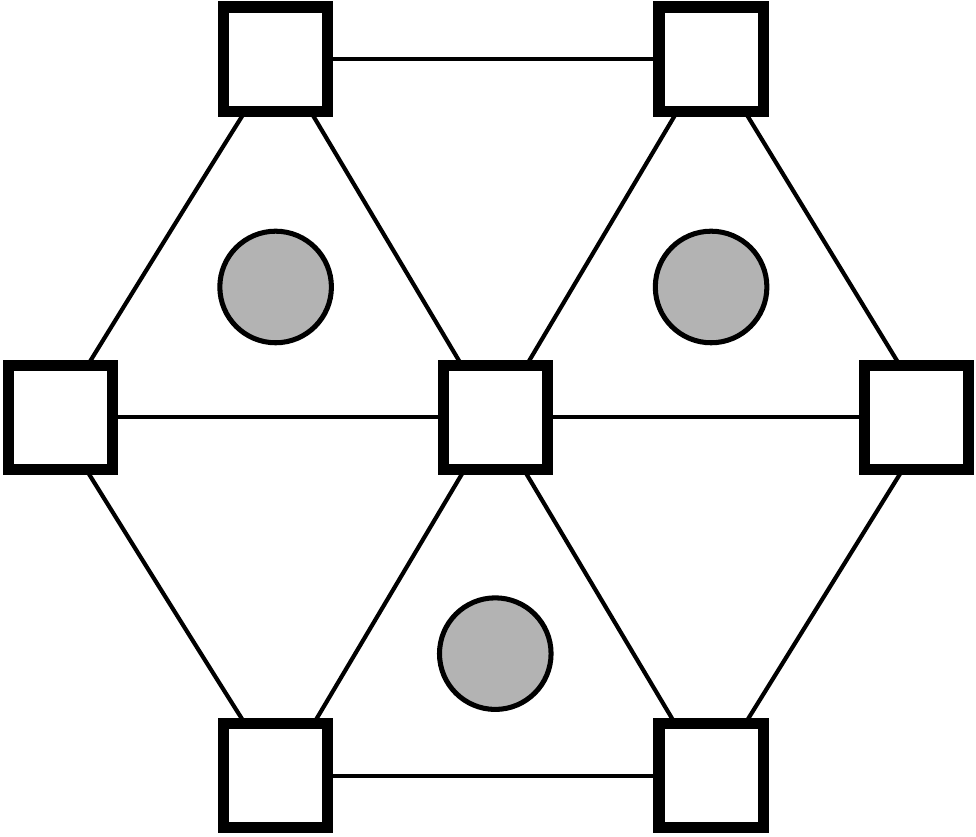} 	&
		\includegraphics[scale=0.16]{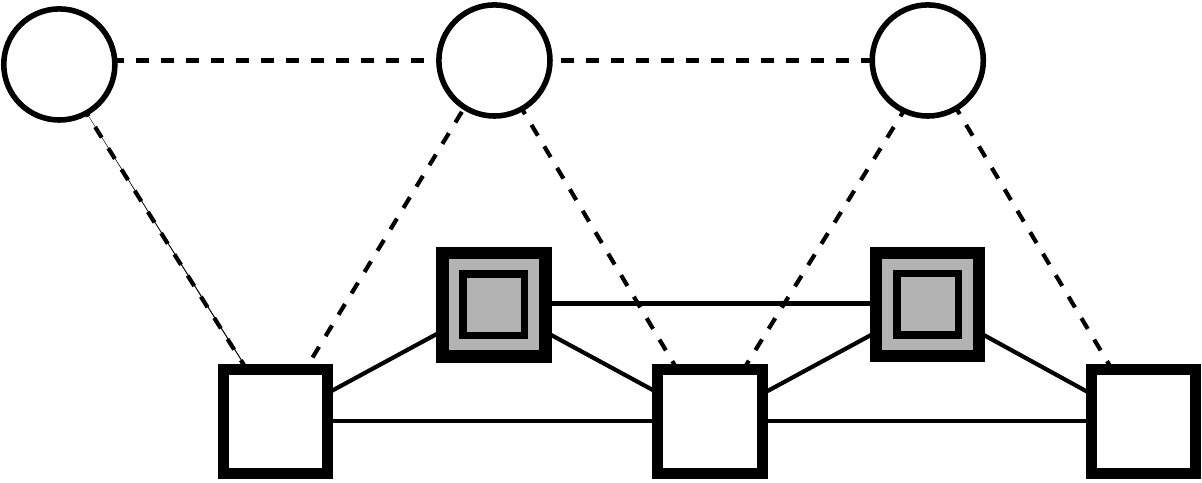}&
		\includegraphics[scale=0.16]{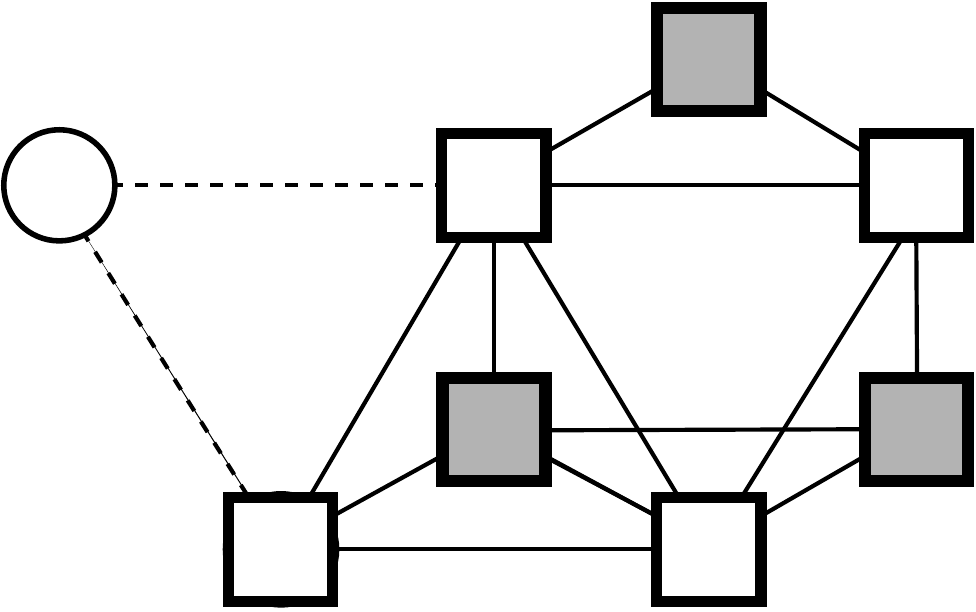}	
		\\
		\hline\hline
		\end{tabular}
	\end{center}
	\caption{Most stable configurations obtained for basal, prismatic and 3D clusters 
	containing between 3 and 7 vacancies. 
	The same conventions as in 
	Tab.~\ref{Tab:additiv_VV} are used to represent atoms and vacancies.}
	\label{fig:smallClusters}
\end{figure}

As compact clusters are the most stable ones, 
we can now separate these clusters in different groups so as to compare their relative stability.
We consider three different groups:
\begin{itemize}
	\item basal clusters, where all vacancies are lying in the same basal plane. 
		These clusters can be seen as precursors of $\langle c \rangle$ loops.
	\item prismatic clusters, where all vacancies are lying in the same prismatic corrugated plane
		(plane denoted A$\alpha$ in Fig. \ref{fig:prismFault}a).
		These clusters can be seen as precursors of $\langle a \rangle$ loops.
	\item 3D clusters maximizing the number of vacancies in position of first nearest-neighbors (precursors of cavities).
\end{itemize}
We investigate different configurations for each group and retain only the most stable ones.
The configurations obtained for clusters containing between 3 and 7 vacancies are shown in Fig.~\ref{fig:smallClusters}.
The most stable configuration in a given group is always the one for which the number of first nearest-neighbor vacancies is maximum.

\begin{figure}[!bth]
	\begin{center}
		\includegraphics[width=0.9\linewidth]{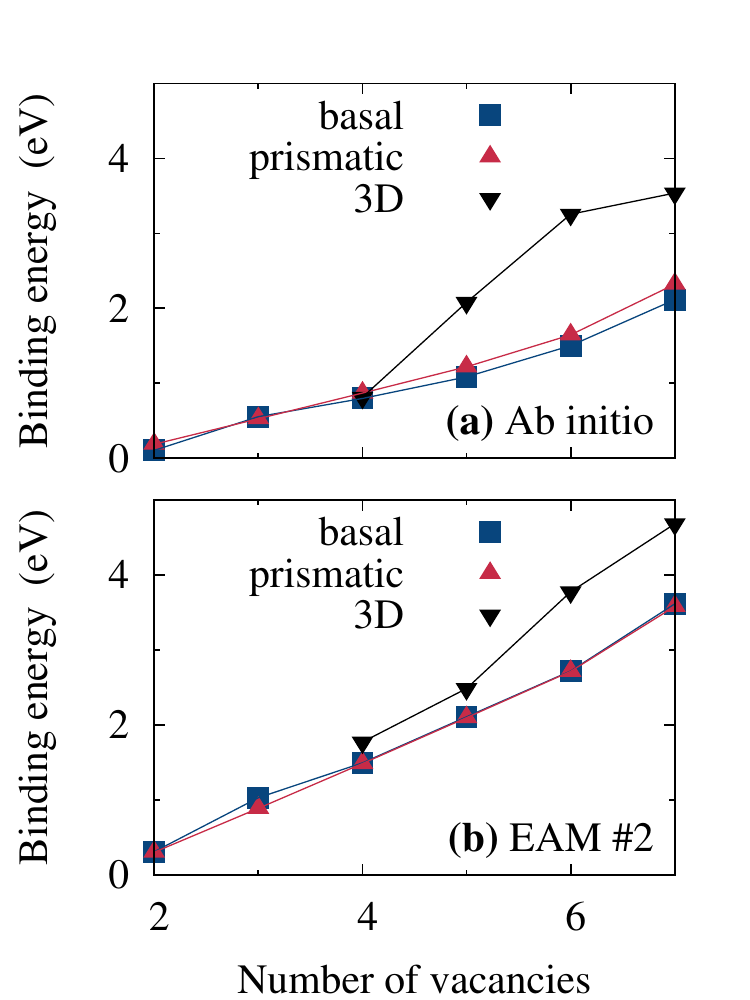}
	\end{center}
\caption{Total binding energies for the different types of vacancy clusters (basal, prismatic and 3D) 
calculated with different energy models: (a) \abinitio calculations and (b) EAM \#2.}
\label{fig:comp_f_amas}
\end{figure}

The binding energies for the different types of clusters containing between 2 and 7 vacancies are shown in  Fig.~\ref{fig:comp_f_amas}.
Both \abinitio calculations and the EAM \#2 potential show that the 3D clusters are the most stable. 
This is not surprising as these clusters are the most compact and therefore maximize the number of attractive interaction between 
vacancies in first nearest-neighbor positions.
Like for the di--vacancy, \abinitio calculations show that the prismatic clusters are slightly more stable 
than the basal ones, for clusters containing at least 5 vacancies.
The empirical potential, on the other hand, predicts the same stability for both types of plane clusters.
Despite this limitation, and an overestimation of cluster stability, 
this empirical potential manages to give a reasonable description of vacancy clusters. 
In particular, the configurations of the most stable clusters predicted by this potential 
are the same as the \abinitio ones for each cluster type, 
and the defect structures after atomic relaxation is also equivalent. 
The EAM \#2 potential appears therefore well-suited to study vacancy clustering in hcp Zr.

\section{Stacking faults and surfaces}

\begin{table*}[bthp]
\caption{Surface energies, $\sigma_{0001}$, $\sigma_{10\bar{1}0}$, and $\sigma_{10\bar{1}1}$,
and stacking fault energies, $\gamma_{E}$, $\gamma_{I_2}$, $\gamma_{I_1}$, and $\gamma_{10\bar{1}0}$,
in hcp Zr. 
The results of the present work, calculated either with \pwscf or the empirical potentials 
EAM \#2 and \#3 are compared to other \abinitio values from the literature obtained with \vasp 
\cite{Domain2004a,Udagawa2010,Poty2011}.
All energies are given in mJ.m$^{-2}$. }
\label{Tab:surf_ener}
\begin{center}
\begin{tabular}{lcccccc}
\hline \hline
& \multicolumn{4}{c}{\Abinitio}	& \multicolumn{2}{c}{EAM} \\
&   \pwscf & Ref. \cite{Domain2004a} & Ref. \cite{Udagawa2010} & Ref. \cite{Poty2011} & \#2 & \#3 \\
  \hline
 $\sigma_{0001}$ 	  & 1600  & 1560 & 1600 &  -- & 1270 & 1540 \\
 $\sigma_{10\bar{1}0}$  & 1670  & 1640 & 1660 &  -- & 1340 & 1540 \\ 
 $\sigma_{10\bar{1}1}$  & 1550  &   -- &   -- &  -- & 1340 & 1550 \\
 $\gamma_{E}$             &  274  &  249 &   -- & 300 &  164 & 297  \\
 $\gamma_{I_2}$           &  213  &  200 &  227 & 228 &  110 & 198  \\
 $\gamma_{I_1}$           &  147  &  124 &   -- & 168 &   55 &  99  \\
 $\gamma_{10\bar{1}0}$  &  211  &  145 &  197 &  -- &  357 & 135  \\
\hline \hline
\end{tabular}
\end{center}
\end{table*}

As vacancy clustering leads to faulted dislocation loops,
it is worth looking at stacking fault energies before studying 
the stability of large vacancy clusters. 
We also study energies of different plane surfaces,
as these surface energies will be used then to model cavities.
Comparison between \abinitio calculations and results obtained with 
empirical potentials will give insights on the ability of these potentials
to model large vacancy clusters.

\subsection{Basal stacking faults}

Condensation of vacancies in a basal plane results in
the creation of a dislocation loop 
of Burgers vector $\vec{b}_1=1/2\,[0001]$. 
This corresponds to the removal of a platelet of one atomic layer 
in the perfect stacking $BABABA$ of basal planes 
and leads to the formation of a highly energetic stacking sequence, 
$BAB.BABA$, denoted $BB$ in the following.
The stacking can then evolve so as to lower the energy of the vacancy loop
by creating two different stacking faults \cite{Hull2011}: 
an extrinsic fault E, which corresponds to the stacking $BABCABA$
and does not change the Burgers vector of the dislocation loop, 
or an intrinsic fault I$_1$, which corresponds to the stacking $BABCBCB$
and leads to a dislocation loop with Burgers vector $\vec{b}_2 = 1/6\,\langle 20\bar{2}3 \rangle$.
To better understand the formation and stability of these different stacking faults
we use the concept of generalized stacking faults \cite{Vitek1968,Vitek2008}. 

\subsubsection{Extrinsic stacking fault}

\begin{figure}[!bthp]
	\begin{center}
		(a) \includegraphics[width=0.8\linewidth]{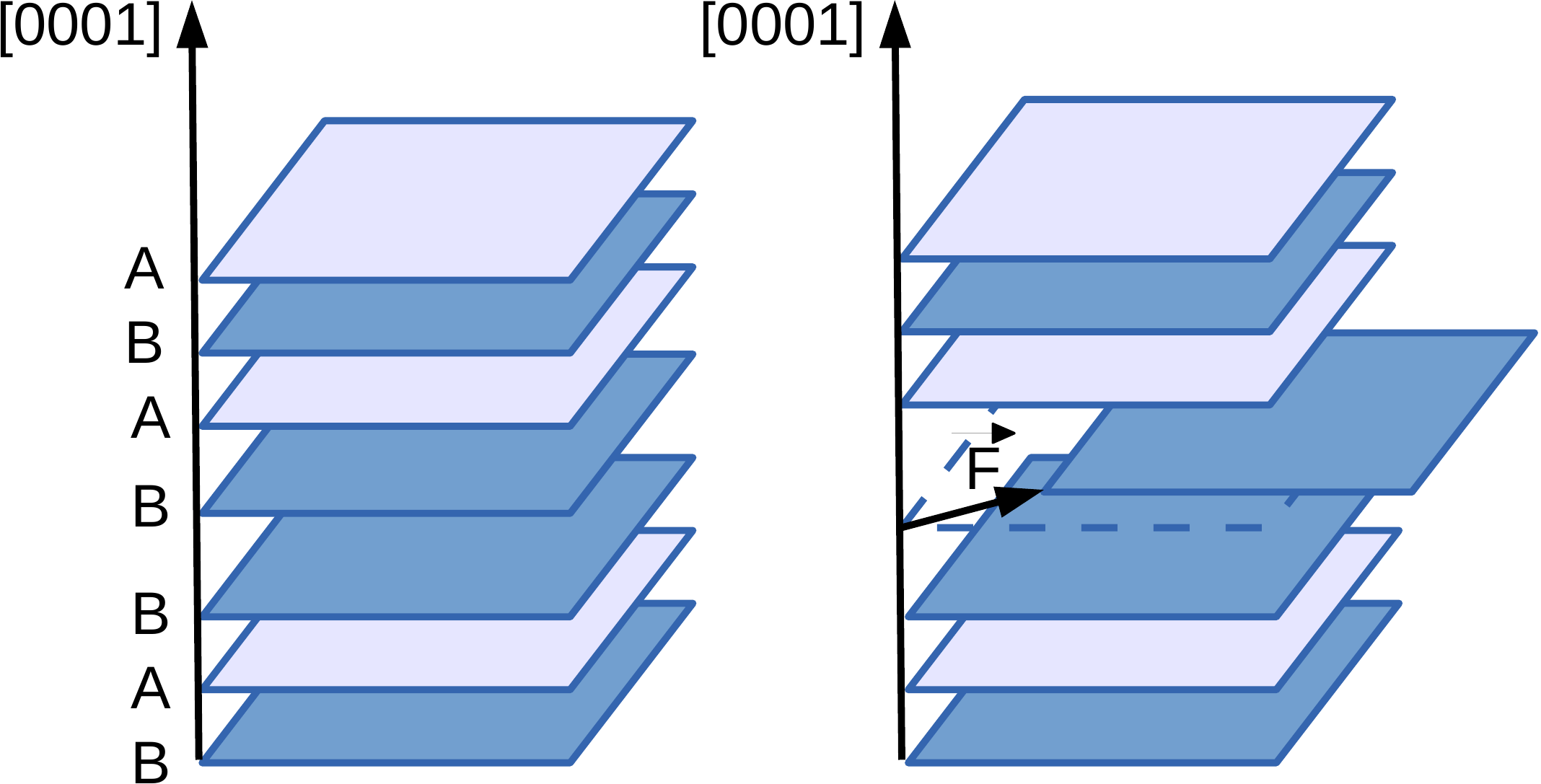} \\
		\vspace{2mm}
		(b) \includegraphics[width=0.8\linewidth]{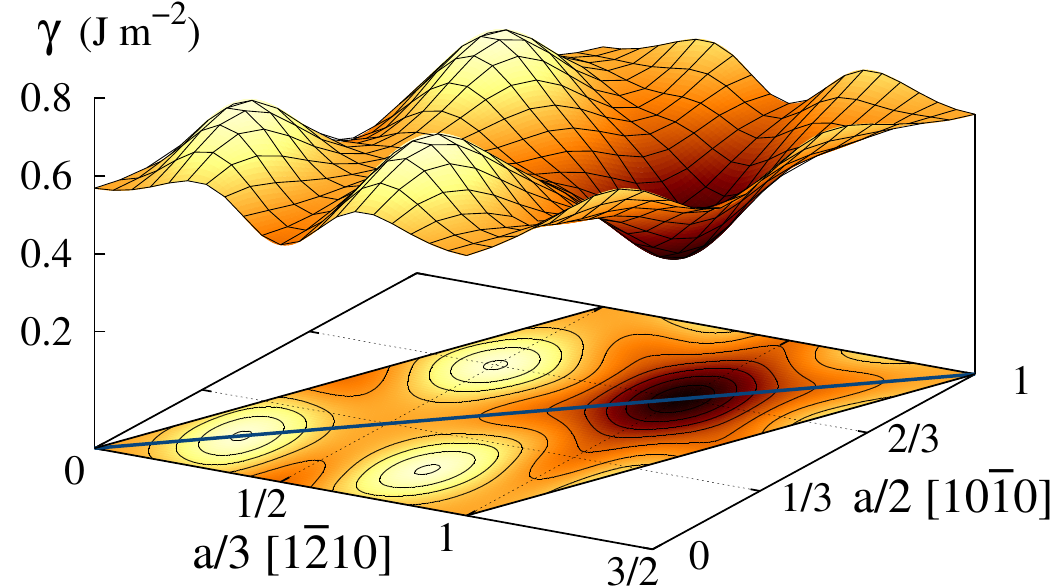} \\
		\vspace{2mm}
		(c) \includegraphics[width=0.8\linewidth]{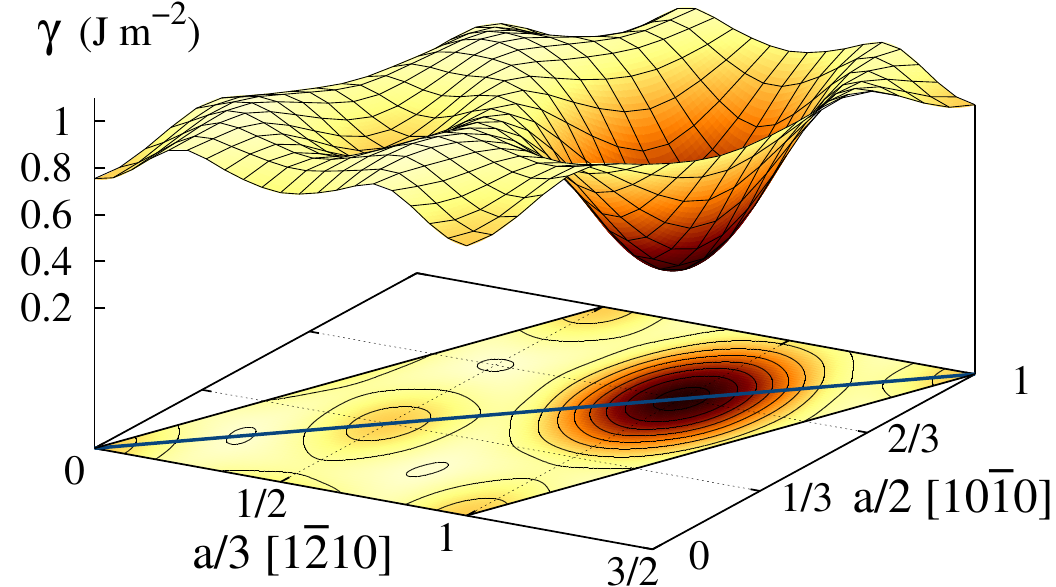} \\
		\vspace{2mm}
		(d) \includegraphics[width=0.8\linewidth]{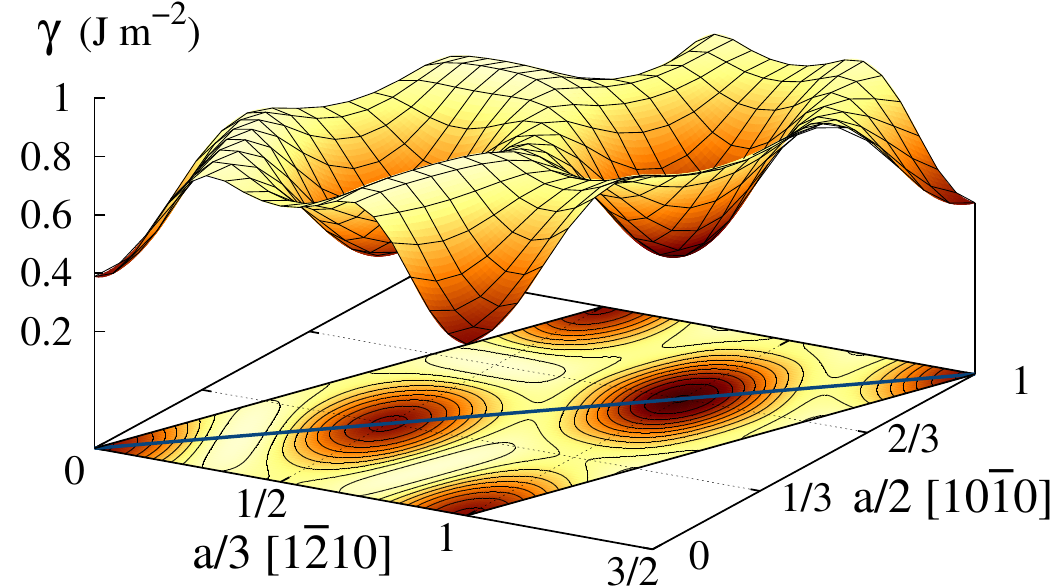} \\
		\vspace{2mm}
		(e) \includegraphics[width=0.8\linewidth]{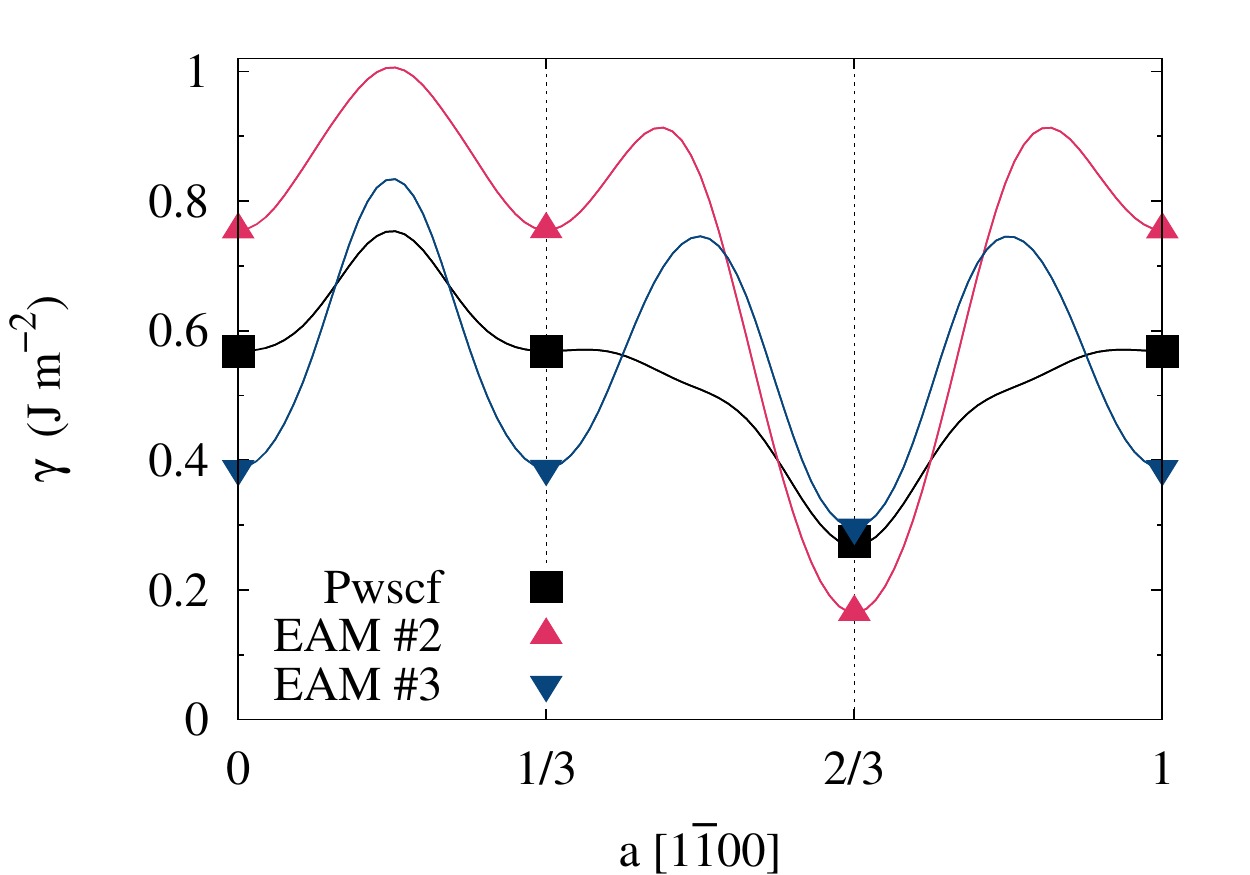}
	\end{center}
	\caption{Extrinsic E generalized stacking fault. 
	(a) Formation mechanism of the  stacking fault.
	(b) \Abinitio, (c) EAM \#2 and (d) EAM \#3 $\gamma$-surfaces.
	(e) Comparison of the fault energies obtained with the different energy models
	along the $[1\bar{1}00]$ direction.}
	\label{fig:basalE}
\end{figure}

The extrinsic fault E is formed from the $BAB.BABA$ stacking by the glide of one atomic plane
(Fig.~\ref{fig:basalE}a). We compute the stacking fault energy for different 
glide vectors $\vec{F}$ lying in the basal plane. In these calculations, 
atoms are allowed to relax only in the direction perpendicular to the fault plane. 
We used a stacking of 15 $(0001)$ planes in the \abinitio calculations, 
which corresponds to a distance $h_{0001}=15c/2$ between fault planes 
and is high enough to prevent any interaction between the fault plane and its periodic images.
Generalized stacking fault energies are calculated on a regular $10\times10$ grid
and are then interpolated with Fourier series.

The obtained energy as a function of the fault vector, or $\gamma$-surface 
(Fig.~\ref{fig:basalE}b, c and d),
shows a minimum for a fault vector $2/3\,[1\bar{1}00]$ which corresponds 
to the metastable extrinsic stacking fault E.
This is the only minimum which exists on the \abinitio $\gamma$-surfaces. 
In particular, the BB stacking, corresponding to a fault vector $\vec{0}$ 
or $1/3\,[1\bar{1}00]$, is unstable.
This is more clearly seen on Fig.~\ref{fig:basalE}e which corresponds to a plot 
of the fault energy along the $[1\bar{1}00]$ direction. 
On the other hand, both empirical potentials EAM \#2 and \#3
predict that the BB stacking is an energy minimum. 
This artifact of empirical potentials leads them to stabilize the BB stacking 
for small vacancy loops, whereas one expects from the \abinitio results 
that such a BB stacking will naturally relax to create an extrinsic E fault.
This may be the reason why special relaxation techniques had to be used 
in Ref.~\cite{Diego2011} to obtain the stable structure of vacancy clusters
lying in the basal planes.

The minimum energy, corresponding to the extrinsic stacking fault, is
$\gamma_E=274$\,mJ.m$^{-2}$, with \abinitio calculations,
in good agreement with already published values \cite{Domain2004a,Poty2011}.
The empirical potential EAM \#2 underestimates this fault energy
whereas a good agreement is obtained with EAM \#3 (Tab.~\ref{Tab:surf_ener}).

\subsubsection{Intrinsic stacking fault}

\begin{figure}[!bthp]
	\begin{center}
		(a) \includegraphics[width=0.79\linewidth]{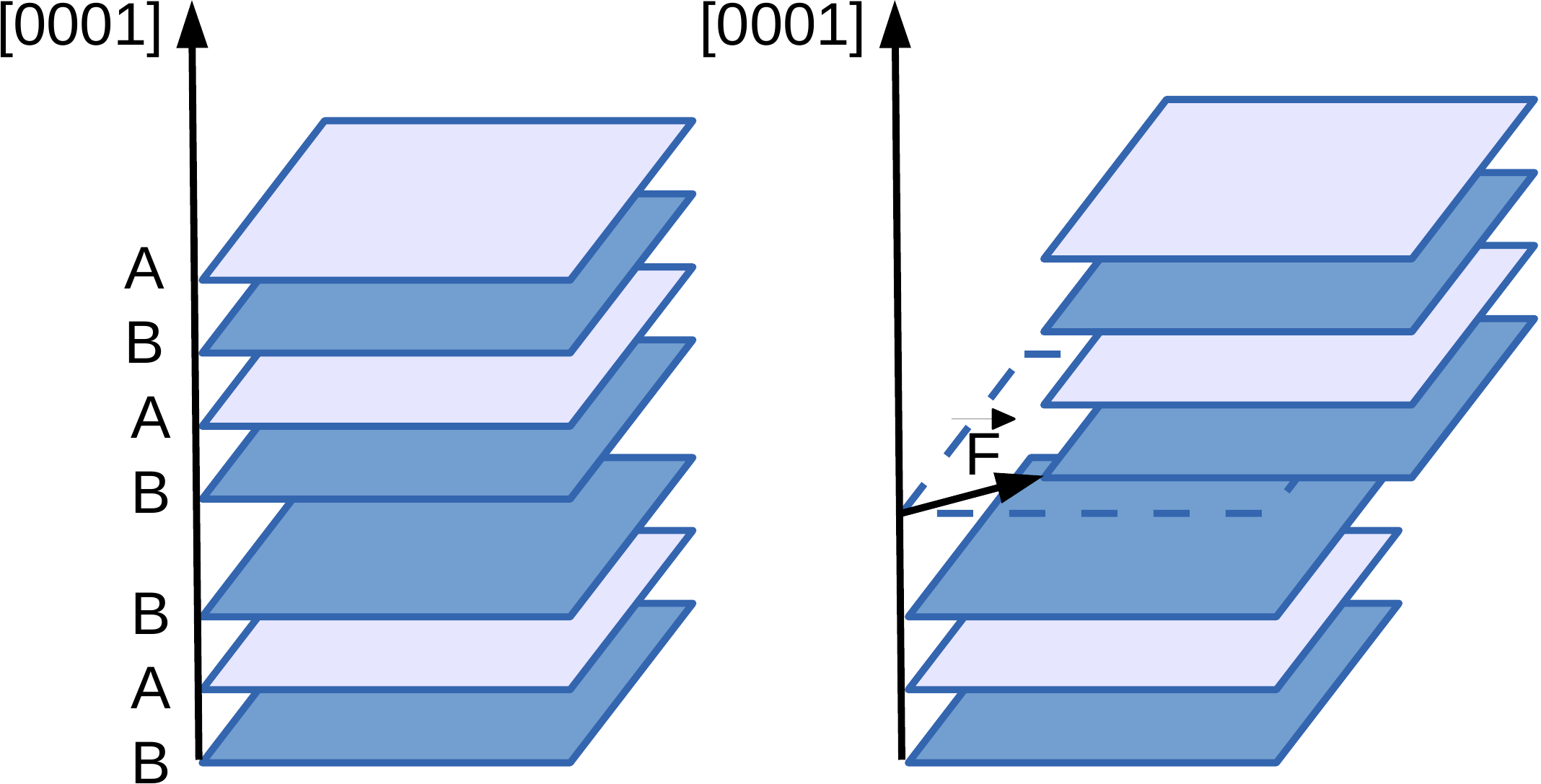} \\
		\vspace{2mm}
		(b) \includegraphics[width=0.79\linewidth]{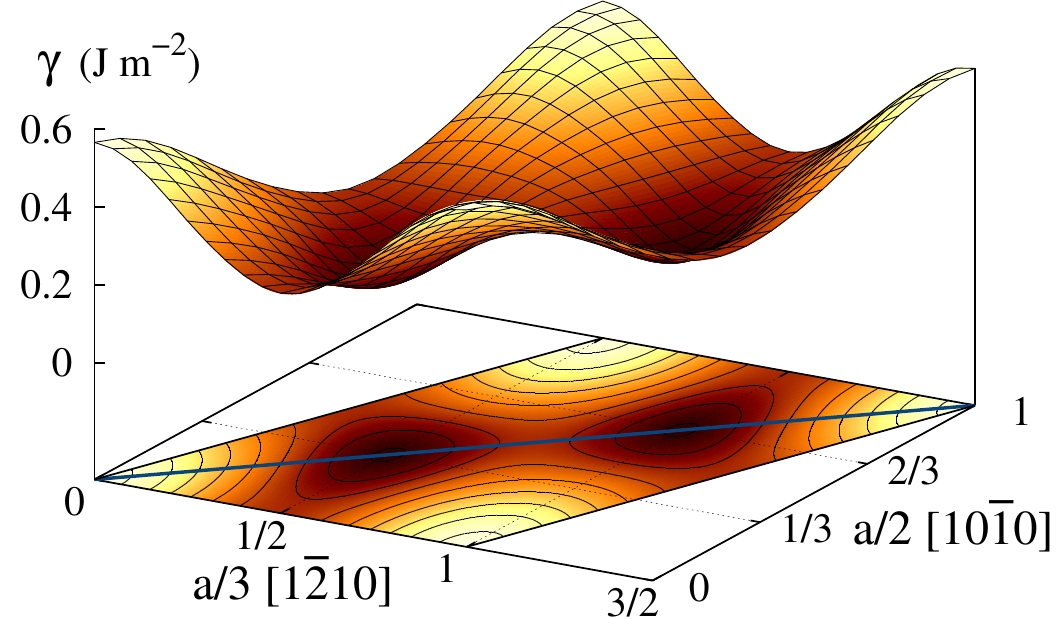} \\
		\vspace{2mm}
		(c) \includegraphics[width=0.79\linewidth]{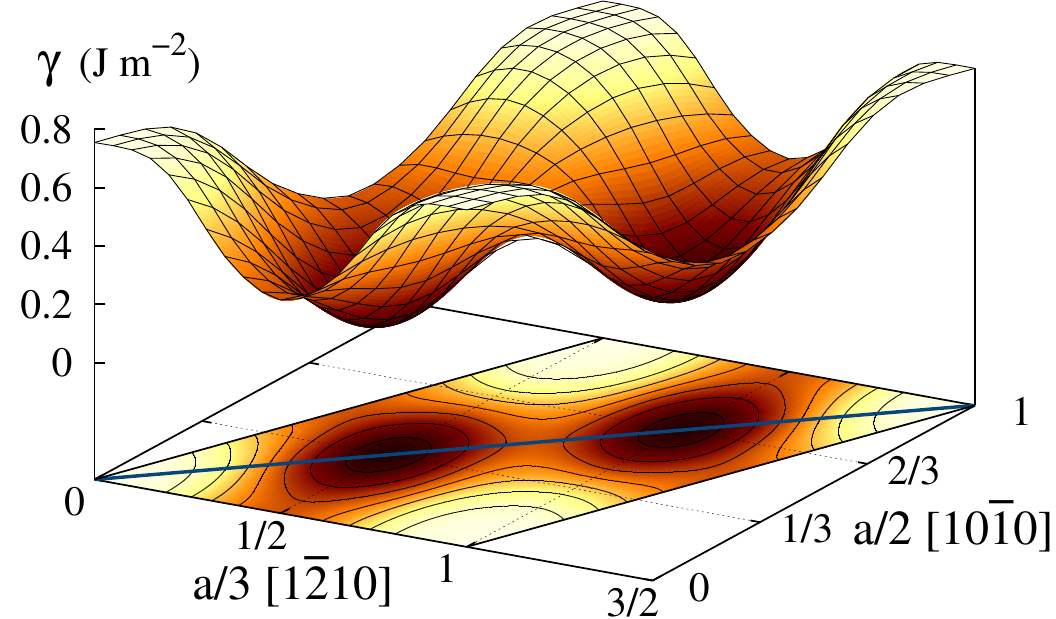} \\
		\vspace{2mm}
		(d) \includegraphics[width=0.79\linewidth]{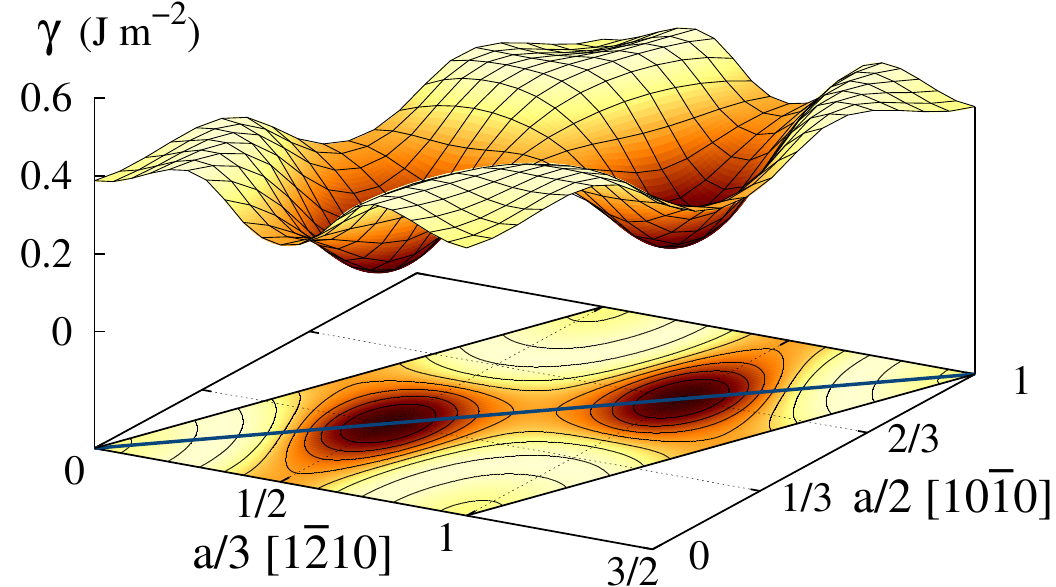} \\
		\vspace{2mm}
		(e) \includegraphics[width=0.79\linewidth]{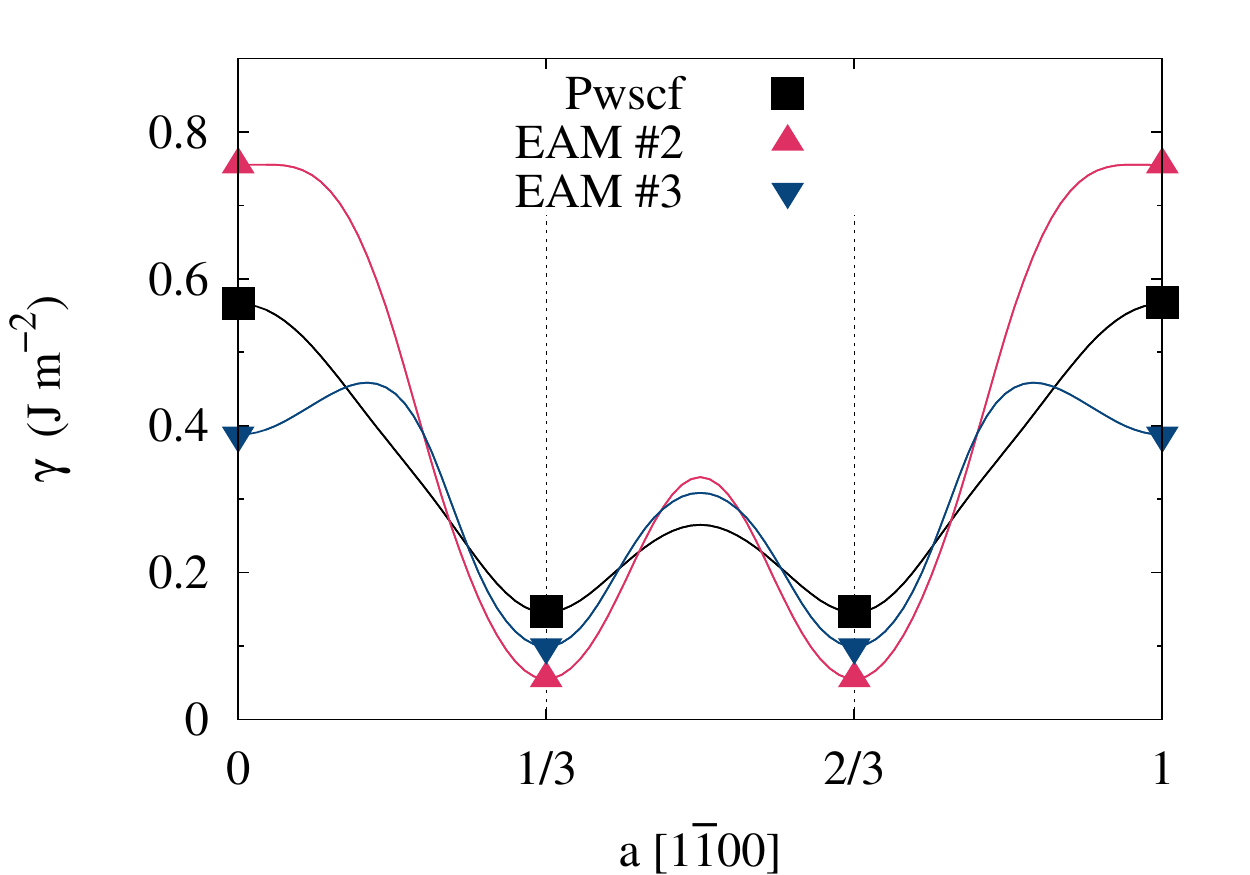} \\
	\end{center}
	\caption{Intrinsic I$_1$ generalized stacking fault. 
	(a) Formation mechanism of the  stacking fault.
	(b) \Abinitio, (c) EAM \#2 and (d) EAM \#3 $\gamma$-surfaces.
	(e) Comparison of the fault energies obtained with the different energy models
	along the $[1\bar{1}00]$ direction.}
	\label{fig:basalI1}
\end{figure}

An intrinsic stacking fault of type I$_1$ is created from the BAB.BABA stacking
by shifting one part of the crystal by a fault vector $\vec{F}$ lying in the basal plane
(Fig. \ref{fig:basalI1}a).
The I$_1$ fault corresponds to a fault vector $1/3\,[1\bar{1}00]$ or $2/3\,[1\bar{1}00]$.
\Abinitio calculations show that this is the only minimum which exists on the corresponding 
$\gamma$ surface (Fig.~\ref{fig:basalI1}b) and that the BB stacking is an energy maximum.
The EAM \#2 potential leads to a $\gamma$ surface in qualitative agreement (Fig.~\ref{fig:basalI1}c). 
On the other hand, EAM \#3 potential predicts that both the BB stacking and the I$_1$ fault 
are energy minima (Fig.~\ref{fig:basalI1}d). Like for the extrinsic generalized stacking fault, this empirical 
potential leads to a stabilization of the BB stacking versus the formation of an intrinsic I$_1$ fault.
It is worth pointing out that a similar artifact had already been mentioned for the basal I$_2$ $\gamma$-surface 
with this potential \cite{Clouet2012}.

\Abinitio calculations lead to an energy minimum $\gamma_{I_1}=147$ mJ.m$^{-2}$, 
still in good agreement with previous \abinitio calculations \cite{Domain2004a,Poty2011}
(Tab.~\ref{Tab:surf_ener}).
Both EAM \#2 and \#3 potentials underestimates this fault energy, 
with the larger error for EAM \#2 (Fig.~\ref{fig:basalI1}e).
All energy models lead to the following order between the energies of the different basal stacking faults:
$\gamma_{I_1} < \gamma_{I_2} < \gamma_E$, in agreement with predictions based on an analysis of  broken bonds
between pairs of atoms \cite{Hirth1982,Hull2011}.
Such an ordering of the fault energies was not retrieved by the long ranged pair potential used in Ref.~\cite{Kapinos1992}.
As EAM \#2 and \#3 are short ranged potentials relying on a central force approximation, they naturally lead 
to the relation $3\gamma_{I_1}=2\gamma_{I_2}=\gamma_E$ \cite{Hirth1982}. 
Our \abinitio calculations show that such a relation is only approximate (Tab.~\ref{Tab:surf_ener}) 
and that the angular contribution of the atomic interaction causes deviations from this idealized picture.
As first pointed out by Legrand \cite{Legrand1984}, a fully predictive modeling of atomic interactions in hcp transition metals 
like Zr needs a proper account of these angular contributions, and thus to go beyond simple empirical potentials 
of the EAM type.

\subsection{Prism stacking faults}

\begin{figure}[!bthp]
	\begin{center}
		\subfigure[]{\includegraphics[height=3.4cm]{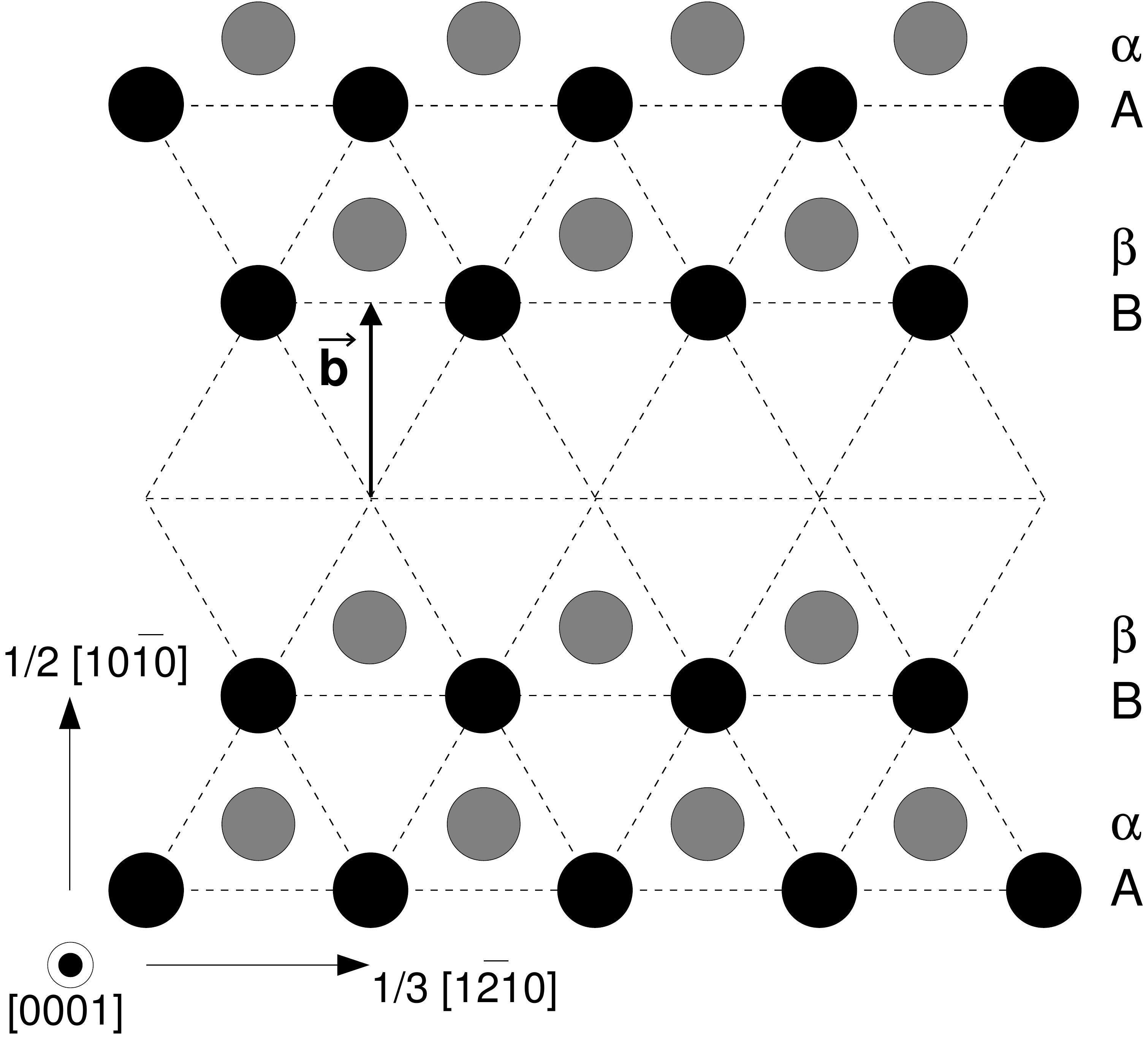}}
		\hfill
		\subfigure[]{\includegraphics[height=3.4cm]{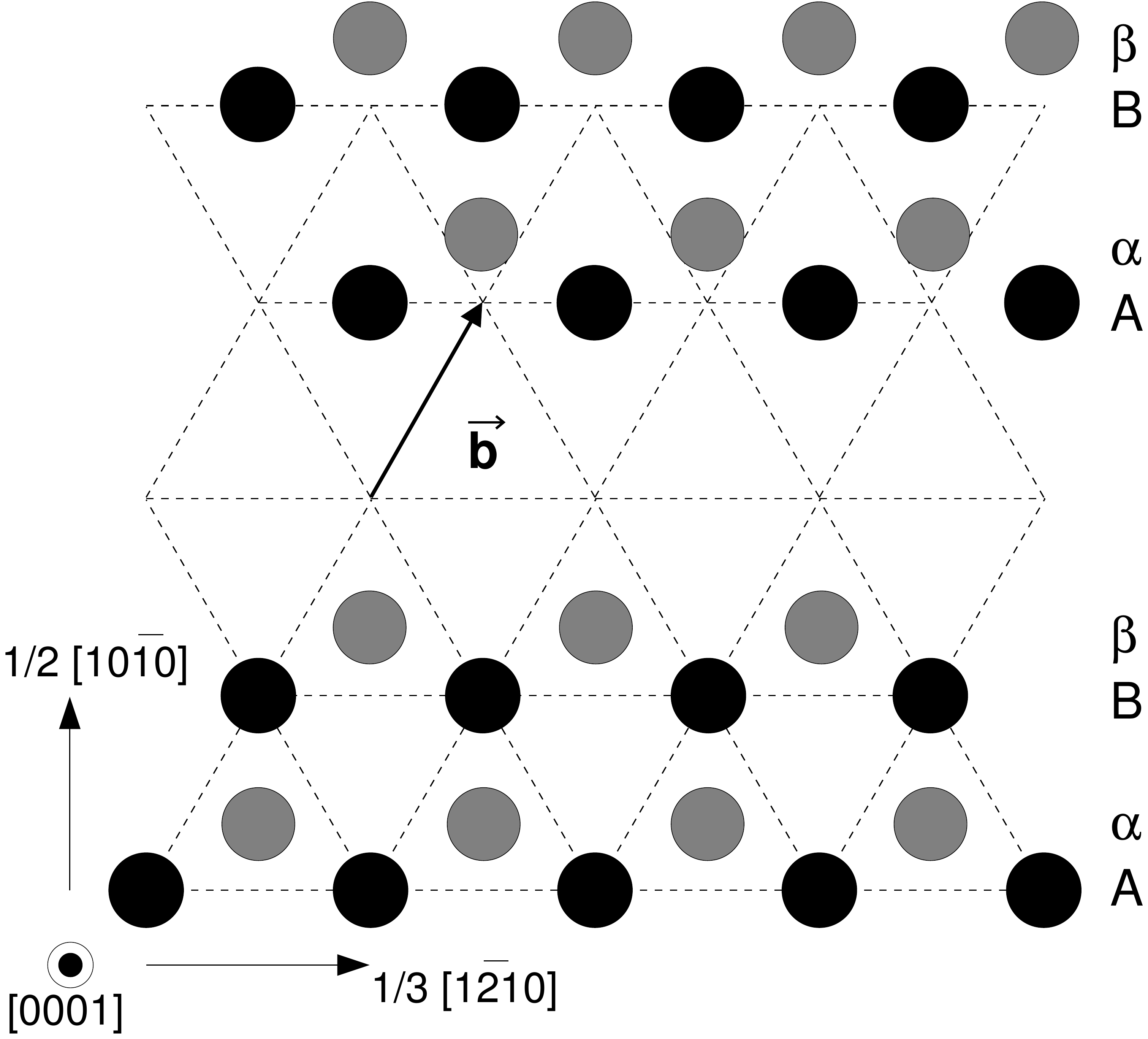}}
	\end{center}
	\caption{Unfaulting mechanism in prismatic planes. 
	(a) The removal of one corrugated $(10\bar{1}0)$ plane creates a prismatic stacking fault.
	(b) The prismatic fault is removed by a shear of amplitude $1/6\ [1\bar{2}10]$ in the $(10\bar{1}0)$ plane.}
	\label{fig:prismFault}
\end{figure}

When removing a vacancy platelet in a corrugated $\{10\bar{1}0\}$ prismatic plane, 
a prismatic stacking fault is formed (Fig.~\ref{fig:prismFault}a). 
This stacking fault, which is associated with the creation of a dislocation loop
of Burgers vector $\vec{b}_3 = 1/2\,\langle 10\bar{1}0\rangle$, 
is the same as the one involved in the dissociation in a prismatic plane
of a $1/3\, \langle 1\bar{2}10 \rangle$ dislocation.
The unfaulting of the vacancy loop occurs by a $1/6\,\langle 1\bar{2}10 \rangle$ shearing of the fault plane,
which leads to a perfect vacancy loop with Burgers vector $\vec{b}_4 = 1/3\,\langle 2\bar{1}\bar{1}0 \rangle$
(Fig.~\ref{fig:prismFault}b).
The $\gamma$-surface associated with this unfaulting mechanism has already been described
in Ref.~\cite{Clouet2012}, both for \abinitio calculations and EAM \#3 potential.
\Abinitio leads to an energy $\gamma_{(10\bar{1}0)}=211$\,mJ.m$^{-2}$
for the metastable stacking fault in this prismatic plane.
EAM \#2 and \#3 respectively overestimates and underestimates 
this fault energy (Tab.~\ref{Tab:surf_ener}).

\subsection{Surface energies}

Large cavities observed by TEM in Zr irradiated with electrons 
show facets in the basal $(0001)$, prismatic $\{10\bar{1}0\}$ 
and pyramidal $\{10\bar{1}1\}$ planes
\cite{Griffiths1993b,Griffiths1994}.
We now consider the surface energies for these three different planes.
For each surface of interest, a crystal block cut in the suitable planes is placed into vacuum. 
For DFT calculations, supercells contain $\sim12$ atomic layers,
and the vacuum slab is $\sim10$\,{\AA} thick.
This ensures the convergence of the surface energies.

Table \ref{Tab:surf_ener} displays the basal, prismatic and pyramidal surface energies,
calculated with the different interaction models.
Our \abinitio calculations show that the pyramidal surface 
has the lowest energy, with the basal surface being the next ones. 
This agrees with TEM observations showing that large cavities 
have facets mainly in the pyramidal and basal planes\cite{Griffiths1993b,Griffiths1994}.
Our \abinitio results for the basal and prismatic surfaces 
match well those of previous \abinitio studies \cite{Domain2004a, Udagawa2010}
(Tab.~\ref{Tab:surf_ener}). 

The EAM \#2 potential underestimates these surface energies 
and predicts that the basal surface is the most stable one,
instead of the pyramidal surface.
Values predicted by EAM \#3 are closer to \abinitio results, 
but this potential does not really discriminate between the different plane surfaces.

\section{Stability of large vacancy clusters: dislocation loops and cavities}

We now study the stability of larger vacancy clusters,  cavities and dislocation loops lying either in the basal or prismatic planes.
As pointed out in the introduction, cavities are hardly observed in irradiated zirconium  \cite{Griffiths1993b,Griffiths1994}
and vacancies mainly condensate in the form of dislocation loops \cite{Northwood1979,Griffiths1987b,Griffiths1988, Onimus2012}.
The formation of $\langle a \rangle$ loops lying in the prismatic planes seems to be more favorable 
than the formation of $\langle c \rangle$ loops in the basal planes. 
$\langle a \rangle$ loops already appear at low irradiation doses, whereas
$\langle c \rangle$ loops only appear for larger irradiation doses, 
when the irradiation growth of the crystal accelerates.
We propose to examine if these experimental observations can be understood through stability arguments. 

\Abinitio calculations cannot be used to study such large clusters. 
On the other hand, the two previous sections have shown
that empirical potentials suffer from limitations. 
We therefore propose to use an hybrid approach to model large vacancy clusters.

We calculate the formation energy of each type of defect for various cluster sizes (up to $\sim 380$ vacancies)
with the EAM \#2 empirical potential.
We choose this potential because it accounts for the binding between vacancies 
and it reasonably describes the relative stability of the vacancy clusters.
(\cf \S \ref{sec:smallClusters}).
Results are then used to validate energy models based on a continuous description of vacancy clusters.
We finally parameterize these continuous models with quantities deduced from \abinitio calculations.
It allows us to extrapolate the DFT results to larger sizes and to discuss the relative stability of the different vacancy defects. 

\subsection{Introduction of vacancy loops in atomistic simulations}

Vacancy loops are introduced in our atomistic simulations by first removing the atoms
inside the vacancy platelets and then applying to all atoms in the simulation box
the displacement field predicted by elasticity theory for the corresponding 
dislocation loop.
The displacement created by a dislocation loop of Burgers vector $\vec{b}$
is given by the Burgers formula \cite{Hirth1982}:
\begin{equation}
	u_i(\vec{x})= -\frac{b_i \Omega(\vec{x})}{4\pi} + \oint_L ...
	\label{eq:Burgers}
\end{equation}
where $\Omega(\vec{x})$ is the solid angle subtended by the loop area at $\vec{x}$.
It corresponds to the plastic displacement created by the loop
and is a purely geometrical term. 
The second term is a closed line integral which accounts for the elastic relaxation.
It can be evaluated using either isotropic \cite{Barnett1985,Barnett2007} or anisotropic \cite{Lazar2013}
elasticity theory.
As Eq. (\ref{eq:Burgers}) is only used to generate the initial configuration,
which is then relaxed with the empirical potential,
we only retain the plastic part of the displacement field.
The solid angle is calculated with the closed-form expression given by Van Oosterom \cite{VanOosterom1983}.

Experimentally \cite{Griffiths1988}, $\langle a \rangle$ loops are circular for radius below $40$\,nm and elliptic above. 
No precise information is available for the shape of the $\langle c \rangle$ loops. 
As these $\langle c \rangle$ loops are formed with a background of numerous $\langle a \rangle$ loops, 
one needs to choose image conditions where $\langle a \rangle$ loops are invisible to see $\langle c \rangle$ loops in TEM. 
As a consequence, these $\langle c \rangle$ loops are usually imaged on their edge.
Dislocation loops are introduced in our simulation boxes as hexagonal loops. 
This morphology is reasonable with regards to the experimental data.
Previous atomistic studies have also shown that the formation energy of the loops only slightly depends on their shape \cite{Diego2008}.

In previous works \cite{Kapinos1992,Kulikov2005,Diego2008,Diego2011}, 
the vacancy loops were obtained by removing vacancy platelets in the relevant planes.
Atomic relaxations, eventually followed by annealing sequences, were used to find the stable configurations.
With the procedure used here, based on the Burgers formula (Eq. \ref{eq:Burgers}),
it is possible to introduce separately each kind of loop 
and to control the stacking fault created by the loop by choosing the corresponding Burgers vector. 

\subsection{Basal dislocation loops}
\label{sec:basalLoops}

We now examine the stability of the different vacancy loops lying in the basal planes.
Loops with a BB stacking fault are formed when choosing a Burgers vector $\vec{b}_1 = 1/2\ [0001]$. 
Loops with an intrinsic fault I$_1$ are formed with $\vec{b}_2 = 1/6\,\langle 20\bar{2}3 \rangle$.
To build loops with an extrinsic fault E, we use the recipe of Hull and Bacon \cite{Hull2011}. 
Two loops of the same size separated by one atomic layer are formed on top of the other 
with Burgers vectors $1/12\,\langle 40\bar{4}3 \rangle$ and  $1/12\,\langle \bar{4}043 \rangle$. 
This results in a loop with Burgers vector $\vec{b}_1 = 1/2\ [0001]$, but with an extrinsic fault.

\begin{figure}
	\begin{center}
	\includegraphics[width=0.9\linewidth]{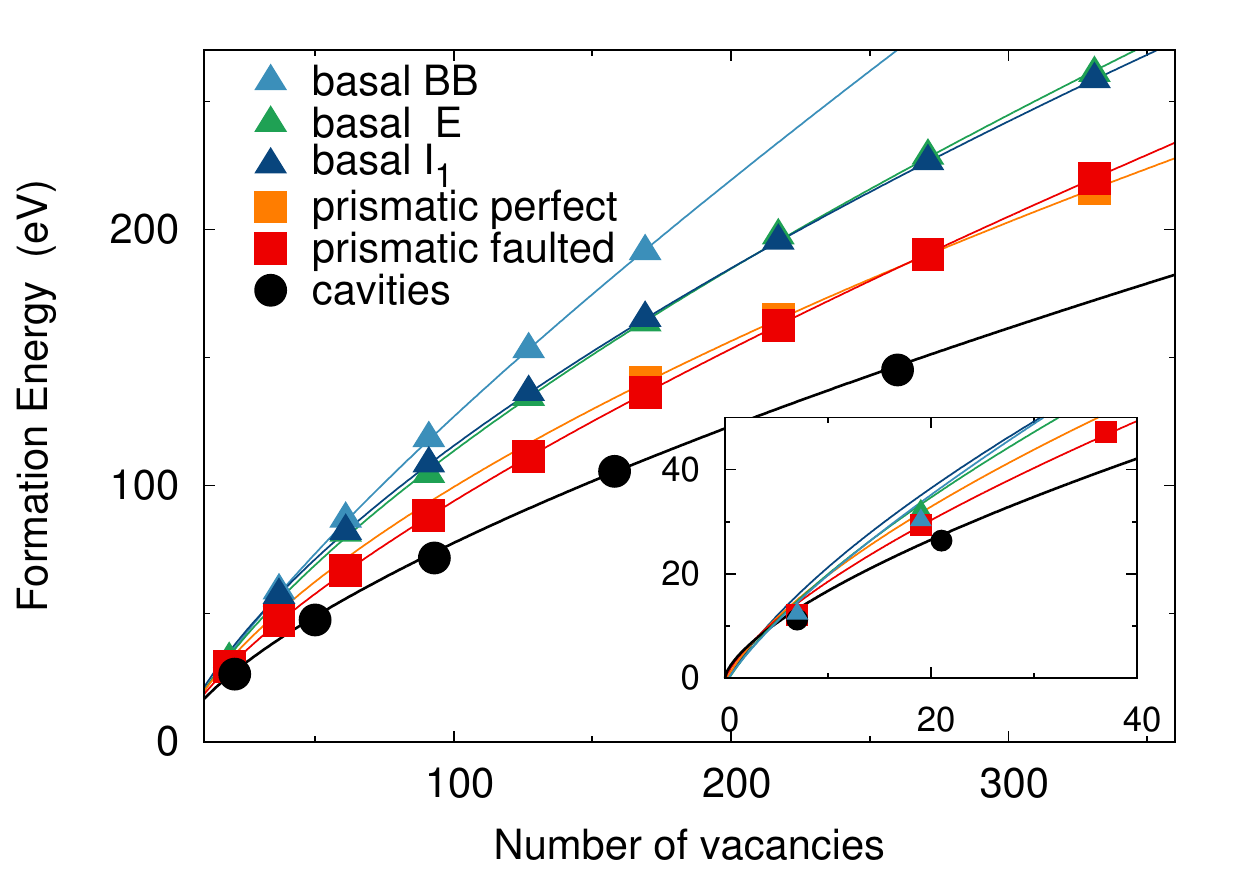}
	\end{center}
\caption{Formation energy of large vacancy clusters in Zr predicted by EAM \#2 potential.
Hexagonal loops lying either in the basal $(0001)$ or the prismatic $\{10\bar{1}0\}$ planes are considered
with different stacking faults, as well as spherical cavities.
The symbols correspond to the results of atomistic simulations
and the lines to continuous laws:
Eq.~\ref{eq:nrj_Ebcle} for dislocation loops 
and Eq.~\ref{eq:nrj_cavity} for cavities.}
\label{fig:Ef_bcles}
\end{figure}

With the EAM \#2 potential, the BB loops are stable for small sizes, 
but they are the less energetically favorable defects (Fig.~\ref{fig:Ef_bcles}).
They become unstable for clusters containing more than 160 vacancies.
The E loops have a lower formation energy than the I$_1$ loops for the small sizes and a greater energy for the largest sizes
(Fig.~\ref{fig:Ef_bcles}).
This is in agreement with a competition between the stacking fault energy ($\gamma_{E} > \gamma_{I_1}$)
and the elastic energy ($\|\vec{b}_1\|^2 < \|\vec{b}_2\|^2$), as it will be seen below.
The stability crossover between both types of loops occurs around 200 vacancies, which corresponds to a loop radius $R=2.4$\,nm.

In order to extrapolate the data for loops containing a larger number of vacancies,
we consider a line tension model \cite{Bacon1980}.
Within this model, the energy of a dislocation loop of radius $R$ is given by:
\begin{equation}
	E^{\rm f}_{\rm loop}(R) = \pi R^2 \gamma + 
		{\oint}_L{ K(\vec{t}) {\ud}s} \ \ln{\left( \frac{R}{r_{\rm c}}\right)},
	\label{eq:nrj_LTmodel}
\end{equation}
where $\gamma$ is the stacking fault energy  
and $r_{\rm c}$ the core radius of the dislocation loop. 
The coefficient $K(\vec{t})$ is the factor appearing in front of the logarithm 
when defining the elastic energy of a straight dislocation. 
It only depends on the bulk elastic constants, the Burgers vector of the loop, 
and the direction of the loop element ${\ud}s$ through its local tangent $\vec{t}$.
We calculate it according to anisotropic elasticity using Stroh sextic formalism \cite{Stroh1958,Stroh1962,Clouet2009a}.
We assume a circular shape to calculate the closed	 line integral 
and define an average value of this coefficient: 
\begin{equation}
	\bar{K} = \frac{1}{2\pi} \oint_0^{2\pi}{ K(\vec{t}) \ud\theta},
	\label{eq:Kprefactor}
\end{equation}
Analytical expressions are available for these dislocation loops lying in the basal plane
(\ref{sec:basalLoops_ani}), but the numerical evaluation has to be done in the more general case, 
in particular for the loops lying in the prismatic plane that will be considered below.
Approximate expressions for this coefficient can also be obtained 
if one assumes isotropic instead of anisotropic elasticity (\ref{sec:isoElas}).
The exact shape of the loop is considered below through a shape factor $f$ used as a fitting parameter.
$f=1$ for a circular loop.
Deviations from this ideal value  occur for non circular loops
because both the perimeter, as a function of the number of vacancies, 
and the average energy coefficient (Eq. \ref{eq:Kprefactor})
depend on the exact shape of the loop.

The link between the loop radius $R$ and the number $n$ of vacancies constituting the loop is established 
considering that $n$ vacancies occupy the same volume as a disk of radius $R$ and of thickness $b_{\rm e}$,
the edge component of the loop Burgers vector normal to its habit plane.
For the three basal loops, we have $b_{\rm e}=c/2$.
Considering that the volume of each vacancy is one atomic volume
$\Omega=\sqrt{3}/4\ a^2c$, the equality $\pi R^2 b_{\rm e} = n \Omega$ 
leads to the relation:
\begin{equation}
	R = a \left(\frac{\sqrt{3}n}{2\pi}\right)^{1/2} = R_1 \sqrt{n},
\label{eq:lien_R_n}
\end{equation}
with $R_1 = a(\sqrt{3}/2\pi)^{1/2}$.
The continuous expression of the basal dislocation loops energy can therefore be expressed as a function of their number $n$ of vacancies: 
\begin{equation}
	E^{\rm f}_{\rm loop} (n)  =  \pi {R_1}^2 \gamma  n 
		+ 2\pi f R_1 \bar{K} \sqrt{n} \ \ln{\left( \frac{R_1 \sqrt{n}}{r_{\rm c}} \right)}.
	\label{eq:nrj_Ebcle}
\end{equation}
As expected, for large loops (large $n$), the stacking fault energy represents the dominant contribution to the formation energy.  
The shape factor $f$ and the core radius $r_{\rm c}$ are used here as fitting parameters, 
in order to obtain the best agreement between the continuous expression $\ref{eq:nrj_Ebcle}$ 
and the results of atomistic simulations.
In this fitting procedure, the stacking fault energies $\gamma$ and the elastic coefficients $\bar{K}$ are fixed to their values
calculated with EAM \#2 for the corresponding basal loop (Tables \ref{Tab:surf_ener} and \ref{Tab:rc_f}).
Despite the simplicity of the line tension model,
the agreement with the atomistic results is good (Fig.~\ref{fig:Ef_bcles}).
The line tension model perfectly fits atomistic simulations for clusters containing at least 50 vacancies.
Some discrepancies appear for smaller clusters (\cf inset in Fig. \ref{fig:Ef_bcles}), 
but the predictions of the line tension model are still reasonable.
The fitted parameters are given in Table~\ref{Tab:rc_f}: 
the shape factor only slightly deviates from its ideal value ($f=1$),
and the core radius $r_{\rm c}$ is close to the norm of a Burgers vector ($b_{\rm e}=c/2$),
as expected from elasticity theory. 
This shows that the integration of the exact shape of the loops through these two fitting parameters
is a reasonable procedure. No attempt was made to calculate them exactly,
as it would require a more complex treatment within elasticity theory 
\cite{Bullough1964,Bacon1970,Schoeck1978}
than the simple line tension model used here.
We also note that
variations of these parameters between the different loops are small.

\begin{table}[htbp]
\caption{Parameters defining the formation energy of the different vacancy loops (Eq.~\ref{eq:nrj_Ebcle}).
The core radii $r_{\rm c}$ (normalized by the lattice parameter $a$)
and the shape factors $f$ have been obtained by fitting the atomistic results of EAM \#2 potential.
The elastic coefficients $\bar{K}$ (in eV/\AA) are deduced from the elastic constants,
corresponding either to EAM  \#2 potential or \abinitio calculations, using anisotropic elasticity 
and Eq.~\ref{eq:Kprefactor}.} 
\label{Tab:rc_f}
\begin{center}
\begin{tabular}{lcccc}
\hline\hline
& $r_{\rm c}$ & $f$ & \multicolumn{2}{c}{$\bar{K}$ } \\
&             &     & EAM & \pwscf \\		
  \hline
 Basal: fault BB               & 0.34  & 1.20 & 0.24  & -- \\ 
 \phantom{Basal:} fault E     & 0.35  & 1.49 & 0.24  & 0.18 \\ 
 \phantom{Basal:} fault I$_1$ & 0.32  & 1.13 & 0.33  & 0.25 \\ 
 Prism: faulted                & 0.11  & 0.85 & 0.22  & 0.20 \\ 
 \phantom{Prism:} perfect     & 0.23  & 1.10 & 0.28  & 0.25 \\ 
\hline\hline
\end{tabular}
\end{center}
\end{table}

\subsection{Prismatic dislocation loops}

We now look at vacancy loops lying in the prismatic $\{ 10\bar{1}0 \}$ planes. 
Faulted loops are created with a Burgers vector $\vec{b}_3 = 1/2\,\langle 10\bar{1}0 \rangle$
and perfect loops with $\vec{b}_4 = 1/3\,\langle 2\bar{1}\bar{1}0 \rangle$.

The stability of these prismatic loops, as predicted by atomistic simulations
using the EAM \# 2 potential,
is given in Fig.~\ref{fig:Ef_bcles}.
Perfect loops are unstable towards the faulted loops when they contain less than 150 vacancies ($R=2$~nm).
In the range where both types of loops are stable, an inversion of stability is observed when increasing the loop size, 
at a size corresponding to $\sim250$ vacancies ($R=2.7$~nm).
For small clusters, faulted loops are the most stable ones because they have a smaller Burgers vector,
whereas perfect loops are more stable for large defects, the stacking fault becoming too costly.
Both types of prismatic loops are more stable than the basal loops. 
This differs from what was obtained in previous atomistic simulations \cite{Kapinos1992,Diego2008,Diego2011},
using different empirical potentials.

We again compare the atomistic results with a continuous law for the dislocation loop energy. 
The expression is still given by Eq.~\ref{eq:nrj_Ebcle}, but with $R_1=\sqrt{ac/2\pi}$ now. 
Using the core radius $r_{\rm c}$ and the shape factors $f$ as fitting parameters,
we obtain a perfect agreement between the continuous laws and the EAM \# 2 energies (Fig.~\ref{fig:Ef_bcles}).
The parameters obtained through this fitting procedure
are given in Table~\ref{Tab:rc_f}:
like for the basal loops, these parameters have reasonable values.

\subsection{Cavities}

We now use the EAM $\#2$ potential to study the stability of cavities. 
They are introduced in simulation boxes as spherical vacancy clusters of increasing size. 
The formation energies are shown in Fig~\ref{fig:Ef_bcles}.
We find that cavities are always more stable than the vacancy-loops, whatever their nature. 
The same result was obtained in Ref.~\cite{Kulikov2005} with a different atomic potential.  

To interpolate these results of atomistic simulations, 
we consider the formation energy of a spherical cavity, 
taking into account only its surface energy. 
This leads for a cavity containing $n$ vacancies to
\begin{equation}
	E^{\rm f}_{\rm cav}(n) = 4 \pi \left( a^2c \frac{3\sqrt{3}}{16\pi}\right)^{2/3}
		f \bar{\sigma} \  n^{2/3}.
	\label{eq:nrj_cavity}
\end{equation}
$f$ is a geometrical factor which is equal to 1 for a spherical cavity. 
It will be taken as a fitting parameter.
The surface energy $\bar{\sigma}$ appearing in this expression is an average energy. 
It can be defined from the energies of plane surfaces, $\sigma_{0001}$, $\sigma_{10\bar{1}0}$
and $\sigma_{10\bar{1}1}$, using the Wulff construction ({\cf} \ref{sec:Wulff}).
This construction ensures that the ideal spherical cavity considered in Eq.~\ref{eq:nrj_cavity}
has the same surface energy as the real faceted cavity.
Using the values predicted by EAM \#2 potential for plane surfaces (Tab.~\ref{Tab:surf_ener}), 
we obtain $\bar{\sigma} = 1420$\,mJ.m$^{-2}$.
The results of atomistic simulations are then perfectly reproduced by Eq.~\ref{eq:nrj_cavity}
with a shape factor $f=1.03$ (Fig. \ref{fig:Ef_bcles}).
The value obtained for this fitting parameter is close
to its ideal value ($f=1$). This shows the validity of our modeling, 
despite its simplicity.

\subsection{\Abinitio modeling}
 
\begin{figure}[bthp]
	\begin{center}
		\includegraphics[width=0.9\linewidth]{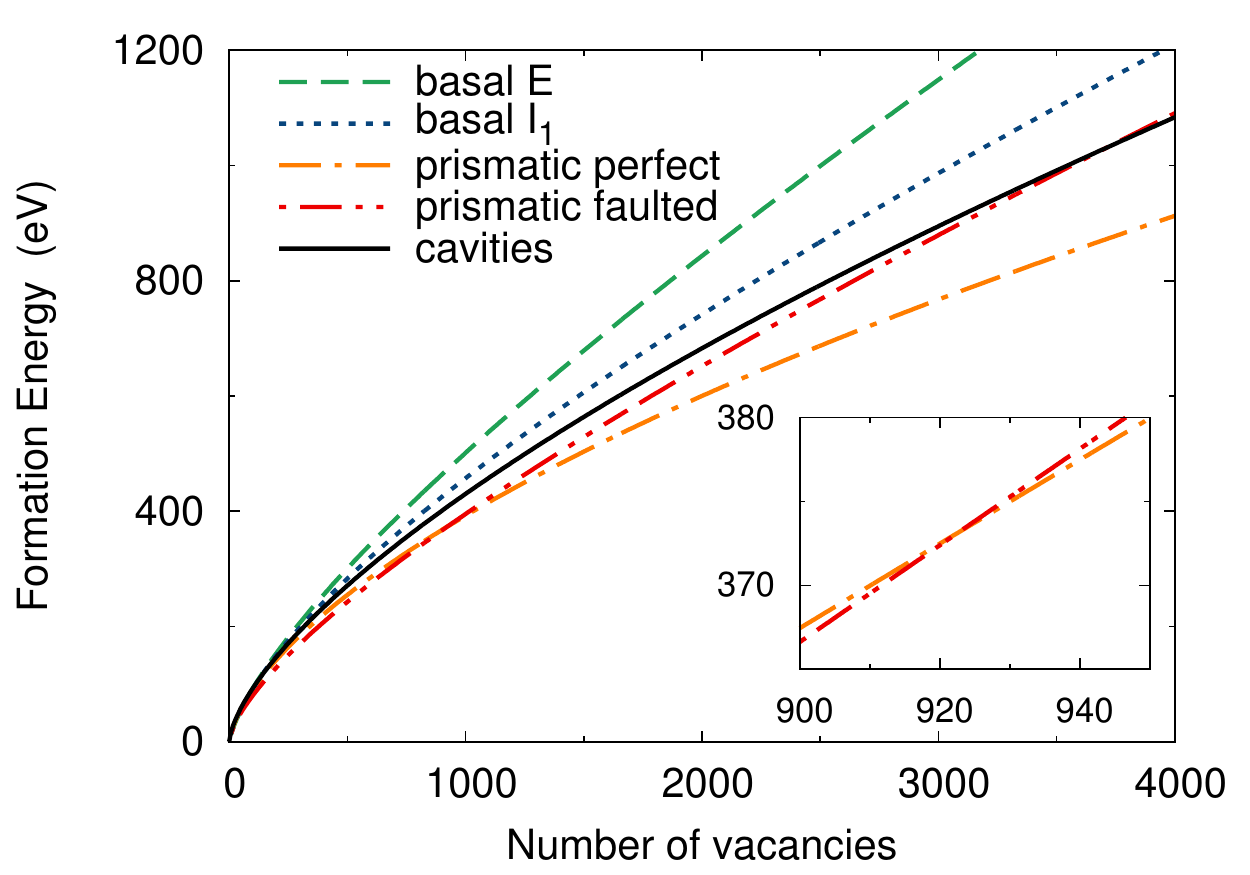}
	\end{center}
	\caption{Formation energies of large vacancy clusters predicted by continuous laws parameterized on DFT results.
	The inset shows the stability inversion between faulted
	and perfect loops lying in the prismatic planes.}
\label{fig:Ef_bclesDFT}
\end{figure}

The continuous laws for the defects energetics, even if their expressions are simple, 
fit well the atomistic results obtained with the EAM \#2 potential. 
We therefore use the same laws, but now with parameters deduced from \abinitio calculations. 
The elastic coefficients $\bar{K}$ and the stacking fault energies 
are fixed to their DFT values.
The surface energy needed for cavities is derived from the same Wulff construction, 
using \abinitio results for plane surfaces. 
This leads to $\bar{\sigma}=1690$\,mJ.m$^{-2}$, a value slightly lower than the experimental 
estimate at 0K, $1900$\,mJ.m$^{-2}$ \cite{Tyson1977}.
The remaining parameters of the continuous models, 
\ie the core radii $r_{\rm c}$ and the shape factors $f$,
are kept fixed to their values deduced from the fitting of EAM \#2 results, 
as these parameters could not be calculated \abinitio.
The resulting stability curves are displayed in Fig.~\ref{fig:Ef_bclesDFT}. 

This \abinitio based modeling predicts that the loops lying in the prismatic planes are the most stable defects
for the whole range of defect sizes. 
This is in agreement with experiments showing that these prismatic $\langle a \rangle$ loops
are the easiest vacancy clusters to create \cite{Onimus2012}. 
For the smallest sizes, the faulted prismatic loops have the lowest formation energy
and when the loop radius becomes larger than $5.2$~nm, perfect prismatic loops become more stable
(\cf inset in Fig. \ref{fig:Ef_bclesDFT}). 

Contrary to what was obtained with the EAM \#2 potential, 
cavities are not the most stable defects.
This disagreement of the empirical potential with \abinitio calculations 
arises both from an underestimation of the surface energy
($\bar{\sigma}=1420$\,mJ.m$^{-2}$ with EAM \#2 instead of 1690\,mJ.m$^{-2}$
with \abinitio) and from an overestimation of the prismatic stacking fault energy
($\gamma_{10\bar{1}0}=357$\,mJ.m$^{-2}$ with EAM \#2 
instead of 211\,mJ.m$^{-2}$).
As cavities are observed only in a few cases in pure Zr \cite{Griffiths1988,Griffiths1993b,Griffiths1994}, 
\abinitio predictions appear more reliable than results of EAM \#2 potential.
According to this \abinitio modeling, cavities are 
the less stable defects also at very small sizes. 
This contradicts our DFT calculations for small clusters containing up to 7 vacancies, 
as these calculations indicate that volume aggregate are more stable than plane clusters 
(Fig.~\ref{fig:comp_f_amas}). 
This illustrates the limitations of the continuous laws used to describe cluster energy. 
Such simple laws are valid only for large enough clusters. 
At small sizes, a full atomic description is needed.

We finally comment on loops lying in the basal planes. 
As \abinitio calculations show that the BB fault is unstable, 
BB loops are not considered.
Like with the EAM \#2 potential, the basal loops are less stable than loops lying in the prismatic planes.
The \abinitio model predicts a stability inversion between the E and I$_1$ basal loops
for $R=1.4$~nm.

\section{Conclusion}

The stability of vacancy clusters in pure hcp zirconium
has been studied using an atomistic modeling approach 
based on both \abinitio calculations and empirical potential.
DFT calculations performed for small vacancy clusters show that the interaction 
between vacancies is attractive only when they are first nearest neighbors.
Such an interaction is partly additive, leading to a higher stability for compact clusters. 
The empirical potential EAM \#2 derived by Mendelev and Ackland \cite{Mendelev2007} 
gives a reasonable description of vacancy clustering.
This potential allows us to study clusters containing up to 380 vacancies,
so as to validate simple analytic laws giving a continuous description 
of the formation energy for the different configurations.
A good agreement has been obtained between these two descriptions,
at an atomic and at a continuous scale.
We could then parameterize such continuous laws with \abinitio calculations.
This shows that the most stable vacancy clusters are dislocation loops,
either perfect or faulted, lying in the prismatic planes.
This is in agreement with experiments where such $\langle a \rangle$ loops
are usually the easier to form.

The continuous laws, which were used to describe the formation energy 
of the different vacancy clusters, are classical models 
which can be found in any metallurgy textbook. 
In particular, Eq. \ref{eq:nrj_Ebcle} for planar clusters
directly derives from dislocation theory \cite{Hirth1982},
within an anisotropic line tension approximation.
We emphasize the usefulness of such laws, which allow discussing the relative stability of different loops,
like we did,
and like it was previously done in 
Ref. \cite{Yoshida1963} for vacancy loops in quenched aluminum
or Ref. \cite{Dudarev2008} for interstitial loops in irradiated iron.

These laws then offer a convenient framework to model the kinetics
of point defect clustering, using for instance cluster dynamics simulations \cite{Kiritani1973}.
As the variation of the energy with the size of the loop differs 
from the one corresponding to 3D clusters, like precipitates, 
the long time evolution leads to a coarsening regime different 
from the usual LSW theory. 
The corresponding coarsening model, 
developed by Kirchner, Burton and Speight \cite{Kirchner1973,Burton1986}, 
has been shown to agree with experimental data \cite{Powell1975,Liu1995,Bonafos1998}.
To develop such a modeling of the kinetic evolution of irradiated Zr, 
it will be necessary however to parameterize the same type of continuous laws 
for interstitial clusters.
Finally, these laws are also a good way to study the influence of impurities and alloying elements
on defect stability, through the modification of stacking fault or surface energies \cite{Domain2004a,Udagawa2011}. 

\appendix

\section{Dislocation loop elastic energy}
\subsection{Isotropic elasticity}
\label{sec:isoElas}

The exact calculation of the elastic coefficient $\bar{K}$ 
appearing in the definition of the loop energy (Eq. \ref{eq:nrj_Ebcle})
can be tough.
We use in this appendix isotropic elasticity 
to obtain simple analytical expressions which can be used as a first approximation.
Within isotropic elasticity, the factors defining the elastic energy 
of an edge or a screw dislocation are respectively
\begin{equation*}
	K_{\rm e} = \frac{\mu}{4\pi(1-\nu)}{b_{\rm e}}^2
	\qquad \textrm{and} \qquad
	K_{\rm s} = \frac{\mu}{4\pi}{b_{\rm s}}^2,
\end{equation*}
where $\mu$ is the shear modulus 
and $\nu$ is Poisson's ratio.
For a specific material, a proper choice has to be done to obtain these average elastic constants.

Basal loops with an extrinsic fault E are pure prismatic loops with $b_{\rm e} = c/2$. 
One simply gets in this case
\begin{equation*}
	\bar{K} = \frac{\mu}{4\pi(1-\nu)} \frac{c^2}{4},
\end{equation*}
whatever the shape of the loop.

Components of the Burgers vector for basal loops with an intrinsic fault I$_1$ 
are varying along the loop. 
Using an angle $\theta$ to define the direction of the loop tangent, 
one can write $b_{\rm e} = c/2 + a \sin{(\theta)} \sqrt{3}/3$
and $b_{\rm s} = a \cos{(\theta)} \sqrt{3}/3$.
The average defined by Eq. \ref{eq:Kprefactor} leads for a circular loop to
\begin{equation*}
	\bar{K} = \frac{\mu}{4\pi(1-\nu)} \left( \frac{1}{4}c^2 + \frac{2-\nu}{6} a^2 \right).
\end{equation*}

Faulted loops lying in the prismatic planes are pure prismatic loops with $b_{\rm e}=a \sqrt{3}/2$,
and thus
\begin{equation*}
	\bar{K} = \frac{\mu}{4\pi(1-\nu)} \frac{3a^2}{4}.
\end{equation*}

Finally, for perfect loops lying in the prismatic planes $b_{\rm e}=a[ \sqrt{3} + \sin{(\theta)} ]/2$,
and $b_{\rm s} = a\sin{(\theta)/2}$. This leads to 
\begin{equation*}
	\bar{K} = \frac{\mu}{4\pi(1-\nu)} \left( 1 - \frac{\nu}{8} \right)a^2.
\end{equation*}

\subsection{Basal loops}
\label{sec:basalLoops_ani}

Thanks to the transverse isotropy of the hexagonal crystal,
one can take full account of the elastic anisotropy
and derive exact expressions of the elastic energy for dislocation loops 
lying in the basal planes \cite{Chou1962}.
When the hcp crystal is oriented with the $x$, $y$, and $z$ axis respectively along the
$[10\bar{1}0]$, $[0001]$, and $[1\bar{2}10]$ directions, 
the Stroh matrix defining the elastic energy of a dislocation lying along the $z$ direction
is diagonal with
\begin{equation*}
  \begin{split}
    K_{11} &= \frac{1}{2\pi} \left( \bar{C}_{11} + C_{13} \right) 
    \sqrt{\frac{C_{44}\left( \bar{C}_{11}-C_{13} \right)}{C_{33}\left( \bar{C}_{11}+C_{13}+2C_{44} \right)}}, \\
    K_{22} &= \sqrt{ \frac{C_{33}}{C_{11}} } K_{11}, \\
    K_{33} &= \frac{1}{2\pi} \sqrt{\frac{1}{2} C_{44}\left( C_{11} - C_{12} \right)},
  \end{split}
\end{equation*}
where $\bar{C}_{11}=\sqrt{C_{11}C_{33}}$.
The elastic coefficient of a basal loop with an extrinsic fault E is then
\begin{equation*}
	\bar{K} = \frac{1}{2} K_{22} \frac{c^2}{4}.
\end{equation*}
For a basal loop with an intrinsic fault I$_1$, one gets
\begin{equation*}
	\bar{K} = \frac{1}{2} K_{22} \frac{c^2}{4}
	+ \frac{1}{4}\left( K_{11} + K_{33} \right)\frac{a^2}{3}.
\end{equation*}
No analytical expression is available for loops lying in the prismatic planes. 
For these loops, one needs either to perform a numerical evaluation, 
like the one of the present study (\S \ref{sec:basalLoops}), 
or to use the previous approximations based on isotropic elasticity.

\section{Wulff construction}
\label{sec:Wulff}

We use the Wulff construction \cite{Christian1975,Porter1992} to define an isotropic surface energy $\bar{\sigma}$
from the surface energies $\sigma_{0001}$, $\sigma_{10\bar{1}0}$, and $\sigma_{10\bar{1}1}$
corresponding respectively to the basal, prismatic and pyramidal planes.
Such a construction predicts that the equilibrium shape of cavities is faceted. 
Considering facets only in the $(0001)$, $\left\{10\bar{1}0\right\}$, and $\left\{10\bar{1}1\right\}$ planes,
the surface of each facet type is proportional to
\begin{eqnarray*}
	\Gamma_{0001}       & = & \frac{\sqrt{3}}{6\gamma^2} 
		\left( 3 \sigmaB - \sqrt{9+12\gamma^2} \sigmaPi  \right)^2 ,\\
	\Gamma_{10\bar{1}0} & = & \frac{4}{3} \sigmaP 
		\left( \sqrt{3+4\gamma^2} \sigmaPi - 2 \gamma \sigmaP \right) ,\\
	\Gamma_{10\bar{1}1} & = & \frac{\sqrt{3+4\gamma^2}}{36\gamma^2} 
		\bigg[ 12 \gamma^2 {\sigmaP}^2 \\
	&& - \left( 3 \sigmaB - \sqrt{9 + 12 \gamma^2} \sigmaPi \right)^2 \bigg],
\end{eqnarray*}
where $\gamma=c/a$.

The isotropic surface energy is obtained by considering a spherical cavity
with the same volume and the same surface energy as the faceted cavity.
This leads to
\begin{equation}
	\bar{\sigma} = \sqrt[3]{\frac{\Gamma_{0001} \sigma_{0001} 
	+ 3 \Gamma_{10\bar{1}0} \sigma_{10\bar{1}0}
	+ 6 \Gamma_{10\bar{1}1} \sigma_{10\bar{1}1} }{2\pi}}.
	\label{eq:Wulff}
\end{equation}

\section*{Acknowledgments}
  This work was performed using HPC resources from GENCI-[CINES/CCRT/IDRIS]
  (Grant 2013-096847). 
  AREVA is acknowledged for financial support.

\section*{References}
\bibliographystyle{elsarticle-num} 
\bibliography{varvenne2014}

\end{document}